\documentclass[final,5p,letterpaper,times,twocolumn]{elsarticle}

\usepackage{amssymb}

\usepackage{amssymb}
\usepackage{graphicx} 
\usepackage{multirow}
\usepackage{booktabs} 
\usepackage{bm}
\usepackage{array} 
\usepackage{appendix}
\usepackage{epstopdf}
\usepackage{bbm}
\usepackage{algorithm}
\usepackage{algorithmicx}
\usepackage{algpseudocode}
\usepackage{hyperref}

\usepackage{amsmath,dsfont}
\usepackage{cases,cleveref}
\crefname{equation}{Eq.}{Eqs.}
\usepackage{harpoon}
\usepackage{amssymb}
\usepackage{mathrsfs}
\usepackage{graphics}
\usepackage{color}
\usepackage{geometry}
\usepackage{epsfig}
\usepackage{diagbox}
\usepackage{mathtools}
\usepackage[final]{changes}
\usepackage{comment}
\usepackage{bbm}
\usepackage{lineno}
\usepackage{cuted}
\usepackage{graphicx}
\def\dd{\mathrm{d}}

\def\p{\partial}

\usepackage{wrapfig}
\usepackage{dsfont}
\usepackage{enumerate}
\usepackage{subfigure}
\usepackage{xcolor}
\usepackage{setspace}

\usepackage{enumitem}
\setitemize{leftmargin=2cm}

\journal{Physica D}

\begin{document}

\begin{frontmatter}

  \title{Overcompensation of transient and permanent death rate
    increases in age-structured models with cannibalistic
    interactions}

  \author[1]{Mingtao Xia}
  \affiliation[1]{organization={Courant Institute of Mathematical Sciences},
    addressline={New York University},
    city={New York},
    postcode={10012-1185},
    state={New York}, 
    country={USA}}
\ead{xiamingtao@nyu.edu}

\author[2]{Xiangting Li}

\affiliation[2]{organization={Department of Computational Medicine},
  addressline={UCLA},
  city={Los Angeles},
  postcode={90095-1766},
  state={CA}, country={USA}}
\ead{xiangting.li@ucla.edu}

\author[2,3]{Tom Chou}
\affiliation[3]{organization={Department of Mathematics},
  addressline={UCLA},
  city={Los Angeles},
  postcode={90095-1555},
  state={CA}, country={USA}}
\ead{tomchou@ucla.edu}


\date{\today}

\begin{abstract}
  There has been renewed interest in understanding the mathematical
  structure of ecological population models that lead to
  overcompensation, the process by which a population recovers to a
  higher level after suffering a permanent increase in predation or
  harvesting. Here, we apply a recently formulated kinetic population
  theory to formally construct an age-structured single-species
  population model that includes a cannibalistic interaction in which
  older individuals prey on younger ones.  Depending on the
  age-dependent structure of this interaction, our model can exhibit
  transient or steady-state overcompensation of an increased death
  rate as well as oscillations of the total population, both phenomena
  that have been observed in ecological systems.  Analytic and
  numerical analysis of our model reveals sufficient conditions for
  overcompensation and oscillations.  We also show how our structured
  population partial integrodifferential equation (PIDE) model can be
  reduced to coupled ODE models representing piecewise constant
  parameter domains, providing additional mathematical insight into
  the emergence of overcompensation.
\end{abstract}




\end{frontmatter}

\section{Introduction}
Overcompensation, which describes the phenomenon in which the total
population of a species increases after experiencing removal or
culling \citep{schroder2014less}, has become an increasingly important
concept in ecology. This phenomenon, also termed the ``hydra effect,''
states that a population increases in response to an
\textit{increased} death or removal rate
\citep{schroder2014less,abrams2009does,mcintire2018can}.  These
overcompensation effects have been shown to arise in European green
crab \citep{grosholz2021stage}, perch \citep{ohlberger2011stage}
populations, and \textit{Tribolium} beetles
\citep{Hastings_Costantino1987,Hastings1987}.  Overcompensation to
selective harvesting is often seen in many tree
\citep{harvesting_palm_2003} and fish populations
\citep{WEIDEL,ZIPKIN}. Apical dominance in botany
\citep{AD2017,AD2023}, whereby a central stem dominates secondary
stems can also give rise to a type of pruning-induced
overcompensation.

There are multiple hypotheses for the mechanism underlying
overcompensation, including the removal of apical dominance
\citep{aarssen1995hypotheses,wise2008applying,lennartsson2018growing}
in plant stem populations, development of resistance to herbivory
\citep{wise2005beyond} in plant populations, reduction of competition
or cannibalism in animal populations
\citep{grenfell1992overcompensation,grosholz2021stage}, and
stage-specific interactions
\citep{Hastings_Costantino1987,Hastings1987,sorenson2021intra,liz2022stability}
Other attempts to explain overcompensation also often rely on the
interplay between multiple species, including consumer-resource
competition.
%
%
For example, a three-compartment consumer-resource model which tracks
the amount of food, the number of predators, and the food consumption
rate has been used to construct a model exhibiting
``overcompensation'' arising in the form of time-periodic increases
and decreases of the total predator population
\citep{pachepsky2008between}. Extensions of such consumer-resource
models that incorporate intraspecific cannibalism in which adults prey
on juveniles when food is scarce have also been used to demonstrate
overcompensation \citep{de2007food}. Such consumer-resource models are
constructed for animal populations and assume overcompensation arises
when resources are abundant.

However, a recent biological/experimental report suggested that
overcompensation can arise solely from intraspecific interactions,
especially cannibalism
\citep{Hastings_Costantino1987,Hastings1987,sorenson2021intra,
  TAKASHINA_FIKSEN,grosholz2021stage}.  This motivates us to provide a
mathematical characterization of cannibalism-induced
overcompensation. Single-species discrete-time stage-specific models
have been proposed and shown to exhibit overcompensation as well as
periodic and even chaotic dynamics
\citep{liz2012hydra,liz2022stability}. Recently,
\citep{sorenson2021intra} developed a continuous-time version of these
models based on prior stage-specific models
\citep{frauenthal1983some,diekmann1986simple,hastings1991oscillations}. These
models have been shown to exhibit nontrivial sustained oscillations
through a Hopf bifurcation. But whether and how overcompensation may
arise from such age-specific interactions are simply characterized.

Here, we generalize stage-specific models
\citep{frauenthal1983some,diekmann1986simple,hastings1991oscillations}
by formulating a simple age-structured partial integrodifferential
equation (PIDE) model with a general cannibalistic interaction that
allows us to more formally study overcompensation.  Our structured
PIDE model will be developed from a high-dimensional
kinetic/stochastic theory of age-structured cannibalistic
interactions, which can formally be projected onto an age-structured
logistic-growth-type PIDE model.
%
%
Although continuous time and continuous-age PIDE models have been
proposed \citep{Hastings_Costantino1987,Hastings1987}, they have been
lumped into discrete age bins before analysis. Although population
oscillations can arise in these lumped models, they have not been
analyzed in the context of overcompensation.

Our generalized PIDE model is readily solved numerically, allowing us
to evaluate both its dynamics and how oscillations and
overcompensation, transient or permanent, arise. Distinct from
previous consumer-resource models
\citep{pachepsky2008between,de2007food,van1988cannibalism,
  henson1997cannibalism}, we show that our PIDE model can exhibit a
rich variety of overcompensating dynamics can arise from intraspecific
interactions alone, without being triggered by external factors such
as an increase in resources. Mathematically, logistic-type
discrete-stage self-inhibition models
\citep{liu1987equilibrium,kozlov2017large} have been shown to also
give rise to undamped population oscillations.




%

Besides analyzing our age-structured PIDE model, we also reduce it to
a set of coupled ODEs that more closely resemble multispecies or
multistage ecological population models.  We will discuss and compare
the qualitative differences between an age-structured model and a
stage-structured model in the overcompensation setting. For example,
in \citep{sorenson2021intra}, overcompensation is found to arise in a
simple two-compartment--young and old populations--ODE model. In our
structured population model, we show that a two-compartment ODE
reduction does not admit overcompensation of an increase in death rate
if the birth rates are kept constant, but that three or more
compartments can.

%
%

In fact, our age-structured interacting model, as well as its
ODE-system approximation, can exhibit rich behavior including dynamic
and permanent overcompensation of increases in the death rate and
the emergence of transient or permanent population oscillations
following the loss of stability of a positive stable point
\citep{boyce1999seasonal}. These dynamics allow us to quantitatively
distinguish \textit{transient overcompensation}, where the total
population temporarily increases following a \textit{temporary}
increase in death rate, from permanent, steady-state overcompensation,
in which a permanent increase in death leads to a permanent increase
in the total population.

In the next section, we develop a nonlinear single-species
age-structured model that describes interactions such as cannibalism
in animal populations.
%
Numerical experiments are carried out in Section~\ref{numerical} to
explore conditions under which overcompensation arises and to validate
previous experimental findings. We also explicitly show how our
age-structured PIDE model can be ``discretized'' into systems of ODEs,
allowing us to derive additional corresponding conditions for
overcompensation and oscillating populations. We give concluding
remarks and discuss some future directions in the Summary and
Conclusions section.

\section{Age-structured intraspecies predation model}

Motivated by the above real-world ecological examples, we formally
construct a simple single-species age-structured population PIDE model
for cannibalization that can lead to overcompensation.
%

%
We start with a \textit{linear} kinetic theory framework that was
recently developed to describe the evolution of a probability density
of proliferating cell populations
\citep{aging_pre,chou2016hierarchical,Xia_kinetic}. To track all ages
in a population, we define the vector ${\bf x}_s=(x_1,...,x_s)$ in
which $x_i$ is the age of the $i^{\text{th}}$ individual and $s$ is
the total number of individuals.

We denote the probability that an animal population has $s$
individuals with ages ${\bf x}_s$ at time $t$ to be $\rho_s({\bf x}_s;
t)$. Without loss of generality, we assume that the probability
$\rho_s$ is symmetric in the age variables, \textit{i.e.}, for any
permutation of $(x_1,...,x_s)$ denoted by ${\bf x}_{s}'$, $\rho_s({\bf
  x}_s; t)=\rho_s({\bf x}_{s}'; t)$.  Normalization of $\rho_s({\bf
  x}_s; t)$ also demands $\sum_{s=0}^{\infty} \int_{0}^{\infty}
\rho({\bf x}_s; t)\text{d}{\bf x}_s \equiv 1, \,\, \forall~t$. If we
denote the birth rate and death rate for the $i^{\text{th}}$
individual in the population by $\beta_i$ and $\mu_i$ respectively,
$\rho({\bf x}_s; t)$ satisfies the following PIDE
\citep{chou2016hierarchical}
\begin{equation}
\begin{aligned}
& \frac{\partial \rho_s({\bf x}_{s}; t)}{\partial t}
+ \sum_{i=1}^s\frac{\partial \rho_s({\bf x}_{s}; t)}{\partial x_i}  
=  \\[-2pt]
&\hspace{6mm} -\sum_{i=1}^{s}\!\big(\beta_{i}+\mu_i\big)\rho_{s}({\bf x}_s; t)
+\!\sum_{i=1}^{s+1}\!\int_{0}^{\infty} \!\!\mu_{i}\, 
\rho_{s+1}({\bf x}_{s+1}[x_i=y]; t)\dd{y}\\[4pt]
& \rho_s({\bf x}_s[x_i=0], t) 
\displaystyle = \frac{1}{s} \sum_{j=1}^{s-1} 
\beta_j\,\rho_{s-1}({\bf x}_{s, -i}; t),
\label{kinetic_primitive}
\end{aligned}
\end{equation}
%
where ${\bf x}_{s+1}[x_i=y] \coloneqq (x_{1}, \ldots, x_{i}=y,
x_{i+1},\ldots x_{s+1})$ and ${\bf x}_s[x_i=0] \coloneqq
(x_1,\ldots,x_{i-1},0,x_{i+1},\ldots,x_s)$. Details of the derivation
of the linear kinetic equation ~\eqref{kinetic_primitive} are given in
\citep{aging_pre,chou2016hierarchical,Xia_kinetic}.  The birth and
death rates $\beta_{i}$ and $\mu_{i}$, may depend on the ages of
individuals $x_{j\neq i}$ other than that of the $i^{\text{th}}$
one. Such multi-individual dependences of $\beta_{i}, \mu_{i}$ lead to
correlations and ultimately nonlinear terms.

Here, we assume the birth rate $\beta_{i} = \beta(x_{i},t)$ of
individual $i$ depends on only the age $x_{i}$ of that individual. The
death rate can be decomposed into a natural death rate and a
cannibalistic interaction term, \textit{i.e.},

\begin{equation}
\begin{aligned}
\mu_i=\mu(x_{i}, t) + \sum_{j\neq i} K(x_j, x_i, t),
\end{aligned}
\end{equation}
%
where $\mu(x_{i},t)$ is the natural death rate of individual $i$ and
$K(x_j, x_i, t)$ is the cannibalizing rate of individual $j$ on
individual $i$.  Note that $K(x_j, x_i, t)$ can depend on both the
ages of the predator and the prey, which generalizes the previous
model in \citep{van1988cannibalism} where $K$ only depends on the
prey's age.

With these definitions, the PIDE satisfied by $\rho_s$
becomes
%
\begin{equation}
\begin{aligned}
& \frac{\partial \rho_s({\bf x}_{s}; t)}{\partial t}
  + \!\sum_{i=1}^s\frac{\partial \rho_s({\bf x}_{s}; t)}{\partial x_i}  = \\[-2pt]
& \hspace{8mm} -\!\sum_{i=1}^{s}\bigg[\beta\big(x_i, t\big)
    +\mu(x_i, t)+ \!\sum_{j\neq i}K(x_j, x_i, t) \bigg]\,\rho_{s}({\bf x}_s; t)\\[-2pt]
& \hspace{8mm} + (s+1)\!\int_{0}^{\infty} \!\bigg[\mu(y, t)+ \!\sum_{i=1}^s \!K(x_i, y, t)\bigg]\,
  \rho_{s+1}({\bf x}_{s}, y; t)\dd{y}\\[4pt]
%
%
&\rho_s({\bf x}_s[x_i=0], t) 
\displaystyle =  \frac{1}{s} \sum_{j=1}^{s-1} 
\beta\big(x_j, t\big)\rho_{s-1}({\bf x}_{s, -i}; t),
\label{kinetic}
\end{aligned}
\end{equation}
%
\noindent where ${\bf x}_{s, -i}\coloneqq (x_1,...,x_{i-1}, x_{i+1},...,x_s)$
and the argument $({\bf x}_{s},y)$ indicates an additional $(s+1)^{\rm
  st}$ individual with age $y$.

The population density at age
$x$ can thus be defined as a sum over all possible numbers of
individuals and marginalizing over all but one age:
\begin{equation}
n(x, t)\coloneqq 
\sum_{s=0}^{\infty} \!s \!\int \rho_{s} ({\bf x}_{s}[x_1=x]; t)\dd{\bf x}_{s, -1}.
\label{nxt}
\end{equation}
We now show that the dependence of $K$ on both $x_j$ and $x_i$
generates nonlinear population dynamics that can give rise to
overcompensation of increased death as well as oscillations. Upon
applying the marginalization and summation of Eq.~\eqref{nxt} to
Eq.~\eqref{kinetic}, we obtain a PIDE satisfied by $n(x, t)$:

\begin{equation}
\begin{aligned}
 \frac{\partial n(x,t)}{\partial t}+ & 
\frac{\partial n(x,t)}{\partial x}  \\
\: & = -\mu(x,t)n(x, t) - \!\!\int_{0}^{\infty}\!\! K(x', x, t)n^{(2)}(x', x, t)\dd{x'},\\
n(0, t)  = & \!\int_0^{\infty}\!\!\beta(x, t)n(x, t)\dd{x}.
\end{aligned}
\end{equation}
where

\begin{equation}
\begin{aligned}
  n^{(2)}(x', x, t) \coloneqq
  \sum_{s=0}^{\infty}\! s(s-1)\!\int \!\rho_{s} ({\bf x}_{s}[x_1=x', x_2=x]; t)\dd{\bf x}_{s, -2}.
\end{aligned}
\end{equation}
and ${\bf x}_{s, -2}\coloneqq(x_3,...,x_s)$. Specifically, if the
correlation between $x$ and $x'$ is small, and $s \gg 1$, we can
approximate $n^{(2)}(x', x, t)\approx n(x', t)n(x, t)$ and obtain a
closed-form PIDE for $n(x, t)$:

\begin{equation}
\begin{aligned}
  \frac{\partial n(x,t)}{\partial t}+ 
\frac{\partial n(x,t)}{\partial x} &
  = -\bigg[\mu(x,t)+ \!\int_{0}^{\infty}\!\!\! K(x', x, t)n(x', t)\dd{x'}\bigg]n(x, t),\\
n(0, t) & = \! \int_0^{\infty}\!\!\beta(x, t)n(x, t)\dd{x}.
\end{aligned}
\label{LVmodel}
\end{equation}
Eq.~\eqref{LVmodel} is the most general form of a simple deterministic
model that incorporates a continuously distributed predator-prey
interaction within an age-structured population model
\citep{lotka2002contribution,volterra1928variations}.  Here, the
quadratic interaction term couples predator and prey populations
through the predation kernel $K(x',x,t)$. Previous analyses of
\eqref{LVmodel} and related equations used them to model cannibalism
\citep{TAKASHINA_FIKSEN,TAKASHINA}, particularly egg-larvae
interactions of the \textit{Tribolium} beetle
\citep{Hastings_Costantino1987,Hastings1987}; however, these studies
did not derive the equations from the underlying kinetic theory as we
have shown above nor did they analyze overcompensation in response to
increased death as we will in the next section.


If $K(x',x,t)=0$, Eq.~\eqref{LVmodel} reduces to the classical
age-structured McKendrick model, which does not exhibit permanent
overcompensation.  If $K(x',x,t)\coloneqq k(x, t)\delta(x'-x)$, where
$\delta$ is the Dirac delta function, Eq.~\eqref{LVmodel} coincides
with previously studied age-structured growth models
\citep{liu1987equilibrium,kozlov2017large}, reducing to

\begin{equation}
\begin{aligned}
  \frac{\partial n(x,t)}{\partial t} + 
\frac{\partial n(x,t)}{\partial x} &
  = -\Big(\mu(x, t) + k(x, t)n(x, t)\Big) n(x,t),\\
n(0, t) & =\! \int_0^{\infty}\!\!\beta(x, t)n(x, t)\dd{x}.
\end{aligned}
\label{logistic}
\end{equation}
As we will be interested primarily in steady-state overcompensation,
or population transients associated with instantaneous jumps in the
death rate, we will restrict our analysis to time-independent
$K(x',x)$ and instantaneous changes to otherwise time-independent
$\beta(x)$ and $\mu(x)$.  Dynamically, changing birth and death rates
can be implemented by changing $\beta$ and $\mu$ instantaneously to
new values that subsequently remain constant (time-independent). Thus,
we will henceforth assume time-independent $\beta, \mu$ (and $K$)
after their abrupt change. If a steady-state population density
$n^{*}(x)$ is reached, it will then satisfy

\begin{equation}
\begin{aligned}
\displaystyle \frac{\dd n^{*}(x)}{\dd x} &
= -\bigg[\mu(x) + \!\int_{0}^{\infty}\!\! K(x', x)n^{*}(x')\dd{x'}\bigg] n^{*}(x),\\
n^{*}(0) & = \!\int_0^{\infty}\!\!\beta(x) n^{*}(x)\dd{x}.
\end{aligned}
\label{steady_state}
\end{equation}
Under this setup, we will show that for our model to display
steady-state overcompensation associated with increased death rate, an
interaction kernel $K(x', x)$ that varies with both $x'$ and $x$ is
necessary.

Incidentally, we note that our PIDE model Eq.~\eqref{steady_state} can
be simply extended to describe populations over structured variables
$x$ that represent quantities other than age. For example, if $x$
represents organism size instead of age, it may follow a growth law

\begin{equation}
\frac{\text{d}x}{\text{d}t} = g(x, t).
\label{deterministic_growth}
\end{equation}
Using a similar derivation starting from the multiparticle kinetic
theory for the probability density, we obtain an equation for the
population density $n(x, t)$ that is similar to Eq.~\eqref{LVmodel}

\begin{equation}
\begin{aligned}
\frac{\partial n(x,t)}{\partial t}+ &
\frac{\partial (g(x, t)n(x,t))}{\partial x} \\
\: &   = -\bigg[\mu(x,t)+ \!\int_{0}^{\infty}\!\! K(x', x, t)n(x', t)\dd{x'}\bigg] n(x, t),\\
g(0, t) n(0, t)  & = \!\int_0^{\infty}\!\!\beta(x, t)n(x, t)\dd{x}.
\end{aligned}
\label{LVmodel_new}
\end{equation}
Thus, the dynamics of a
population structured according to variables such as size or weight can be
analyzed using Eq.~\eqref{LVmodel_new}. 

However, if $\tfrac{\text{d}x}{\text{d}t} = g(x)$ (time-inhomogeneous
growth), then by defining $y = \int^x_0 \frac{1}{g(z)}\text{d}z$,
$\text{d}y = \text{d}t$, and the new structured variable $y$ could be
seen as an age.  In this scenario, we shall obtain a differential
equation for $n(y, t)$

\begin{equation}
\begin{aligned}
  \frac{\partial n(y,t)}{\partial t} +
  \frac{\partial n(y,t)}{\partial y} & = -\bigg[\mu(y,t)+\!\int_{0}^{\infty}\!\! K(y', y, t)n(y', t')\dd{y}\bigg]n(y, t),\\
  n(0, t) & = \!\int_0^{\infty}\!\!\beta(y, t)n(y, t)\dd{y}.
\end{aligned}
\label{LVmodel_new0}
\end{equation}
which is identical in form to Eq.~\eqref{LVmodel}. For simplicity and
without loss of generality, we henceforth explore overcompensation in
a population structured according to age or according to an attribute
that grows in a time-inhomogeneous manner, allowing us to use
Eq.~\eqref{logistic} or \eqref{LVmodel_new0}.

\section{Results and Discussion}
\label{numerical}

Overcompensation of the total population can be reflected as a
transient increase in the overall population following a transient
increase in $\mu$, as a permanent change in the steady-state
population and/or as a periodically fluctuating population following
permanent increases in the death rate. Although the general conditions
on $K(x', x)$ required for the model to exhibit overcompensation
and/or oscillations cannot be analytically derived, we present several
cases that preclude or allow overcompensation. We also present a
piecewise constant function approximation to convert our PIDE model to
a system of ODEs, further providing mathematical insight into the
dynamical behavior of our model.

\subsection{Specific interactions that preclude overcompensation}

Here, we consider permanent changes in the birth and death rates
$\beta, \mu$ and present simple interactions $K(x',x)$ for which
permanent, steady-state overcompensation can be proven not to arise:

\begin{itemize}
    \item[(\textit{A.1})] $K(x', x)= k(x)\delta(x'-x), k(x)>0$.
      Correspondingly, Eq.~\eqref{steady_state} reduces to $\tfrac{\dd
        n^{*}}{\dd x}= -\mu(x) n^*-k(x)(n^*)^2$,
      $n^{*}(x=0)=\int_{0}^{\infty}\beta(x)n^{*}(x)\dd x$.
     \item[(\textit{A.2})] $K(x', x)=K(x')$ with constant $\beta,
       \mu$.  This interaction is independent of prey age $x$ and the
       resulting model corresponds to an age-structured McKendrick
       model with a modified death rate $\mu\to
       \mu+\int_0^{\infty}K(x')n(x')\dd{x}$ as proposed in
       \citep{kozlov2017large}.
      %
      %
  \item[(\textit{A.3})] $K(x', x) = K(x)$ with constant $\beta, \mu$.
    This case corresponds to predators of any age $x'$ preferentially
    cannibalizing prey of age $x$ according to $K(x)$. With this
    interaction kernel, Eq.~\eqref{LVmodel} reduces to a linear,
    self-consistent McKendrick equation, as in (ii), except with a
    modified death rate $\mu\to \mu + K(x)N^{*}$. A uniform
    interaction kernel (constant $K$) is a subcase ($X\to \infty$).
\end{itemize}
Here, $\delta(x)$ is the Dirac delta distribution and $\theta(x>0)=1,
\theta(x\leq 0)=0$ is the Heaviside function. All of these cases admit
simple, unique, nonzero steady states. The corresponding reduced
models of cases (\textit{A.1}), (\textit{A.2}), and (\textit{A.3}) all
admit simple self-consistent solutions. For constant birth and death
rates $\beta$ and $\mu$, we prove in Appendix \ref{NOOC} that
interactions (\textit{A.1}), (\textit{A.2}), and (\textit{A.3}) all
preclude steady-state overcompensation; that is, the total
steady-state population $N^{*} \equiv \int_0^{\infty} n^{*}(x)\dd{x}$,
where $n^*$ is the steady-state solution of Eq.~\eqref{steady_state},
does not increase when $\mu$ increases. Case (\textit{A.1}) indicates
that a more distributed kernel is required for overcompensation. Case
(\textit{A.2}) indicates that variation in predator age $x'$ alone is
insufficient to generate steady-state overcompensation. Case
(\textit{A.1}) represents an interaction kernel that varies only in
prey age $x$ and is also insufficient to generate steady-state
overcompensation. These results imply that steady-state
overcompensation requires $K(x',x)$ that varies to some degree in both
the prey age $x$ and predator age $x'$.

\subsection{Specific interactions that may exhibit overcompensation}

 We have also found two simple forms for $K(x',x)$ that allow
 steady-state overcompensation

\begin{itemize}
\item[(\textit{B.1})] $K(x', x)= k\delta(x-a)\delta(x'-b)$, with
  $\beta > \mu, k>0$ constant.  Under this point-source and point-sink
  interaction, the steady-state equation is $\tfrac{\dd n^{*}}{\dd x}=
  -\big[\mu + kn^{*}(b)\delta(x-a)\big]n^*$, which is solved by
  $n^{*}(x) = n^{*}(x=0)e^{-\mu x -k n^{*}(b)\theta(x-a)}$,
\item[(\textit{B.2})] $K(x', x)=k \delta(x)\theta(x'-b)$ with constant
  $\beta> \mu, k >0$. The interaction describes adults with age $x\geq
  b$ feeding, with rate $k$ on newborns or eggs. The steady-state ODE
  becomes $\tfrac{\dd n^{*}}{\dd x}= -\big[\mu +
    kn^{*}(b)\delta(x-a)\big]n^*$, which is solved by $n^{*}(x) =
  n^{*}(0)e^{-\mu x - k N_{\rm b}}$ with $N_{b}\coloneqq
  \int_{b}^{\infty}\!n^{*}(x)\dd x$.
 
\end{itemize}

Appendix \ref{YESOC} provides detailed calculations of the total
population under models (\textit{B.1}) and (\textit{B.2}). Further
analysis shows the conditions (parameter regimes of $\mu, k, \beta$)
under which steady-state overcompensation, $\partial N^{*}/\partial
\mu > 0$, arises.

\subsection{Existence and uniqueness of the positive steady state}

The specific forms of $K(x',x)$ given above that either preclude or
allow overcompensation provides some mechanistic insight into the
interaction that can give rise to overcompensation.  Roughly, the
interaction kernel $K(x',x)$ should have a positive gradient in the
$x'$ direction and a negative gradient in the $x$ direction. This
asymmetry in $K(x',x)$ leads to sufficient suppression of the older,
predating, population such that a ``killing of the killers'' effect
leads to larger total populations.

Henceforth, we consider a fairly general form for $K(x',x)$ that
incorporates the dependencies on both $x'$ and $x$ that is compatible
with overcompensation:

\begin{equation}
K(x',x) = 0, \,\,\, \forall x \geq X, \,\,\, \mbox{or}\,\,\, x' \leq x.
\label{eq:compact_condition}
\end{equation}
Here, $X$ is an age threshold such that any individual above age $X$
cannot be cannibalized.  In Appendix \ref{UNIQUE}, we prove that given
time-independent $\beta(x), \mu(x)$ our model (Eqs.~\eqref{LVmodel} and
\eqref{steady_state}) admits one unique steady state $n^{*}$ under some
conditions.  Thus, for a transient perturbation of the birth and death
rates (which eventually return to their constant pre-perturbation
values) permanent overcompensation of the population cannot arise. The
system has no other accessible steady state and the total population
returns to its unique steady-state value, provided it does not vanish
during its transient evolution. However, abrupt, permanent increases
in the death rate may lead to permanent overcompensation as the new
steady state associated with higher $\mu$ may be associated with a
higher total population $N^{*}=\int_{0}^{\infty}n^{*}(x,t)\dd x$.  In
Appendix \ref{EXISTENCE}, we discuss general
characterization/conditions for existence of the positive steady
state.

\subsection{Overcompensation of increased death rates}

\begin{figure*}
\centering \includegraphics[width=0.97\textwidth]{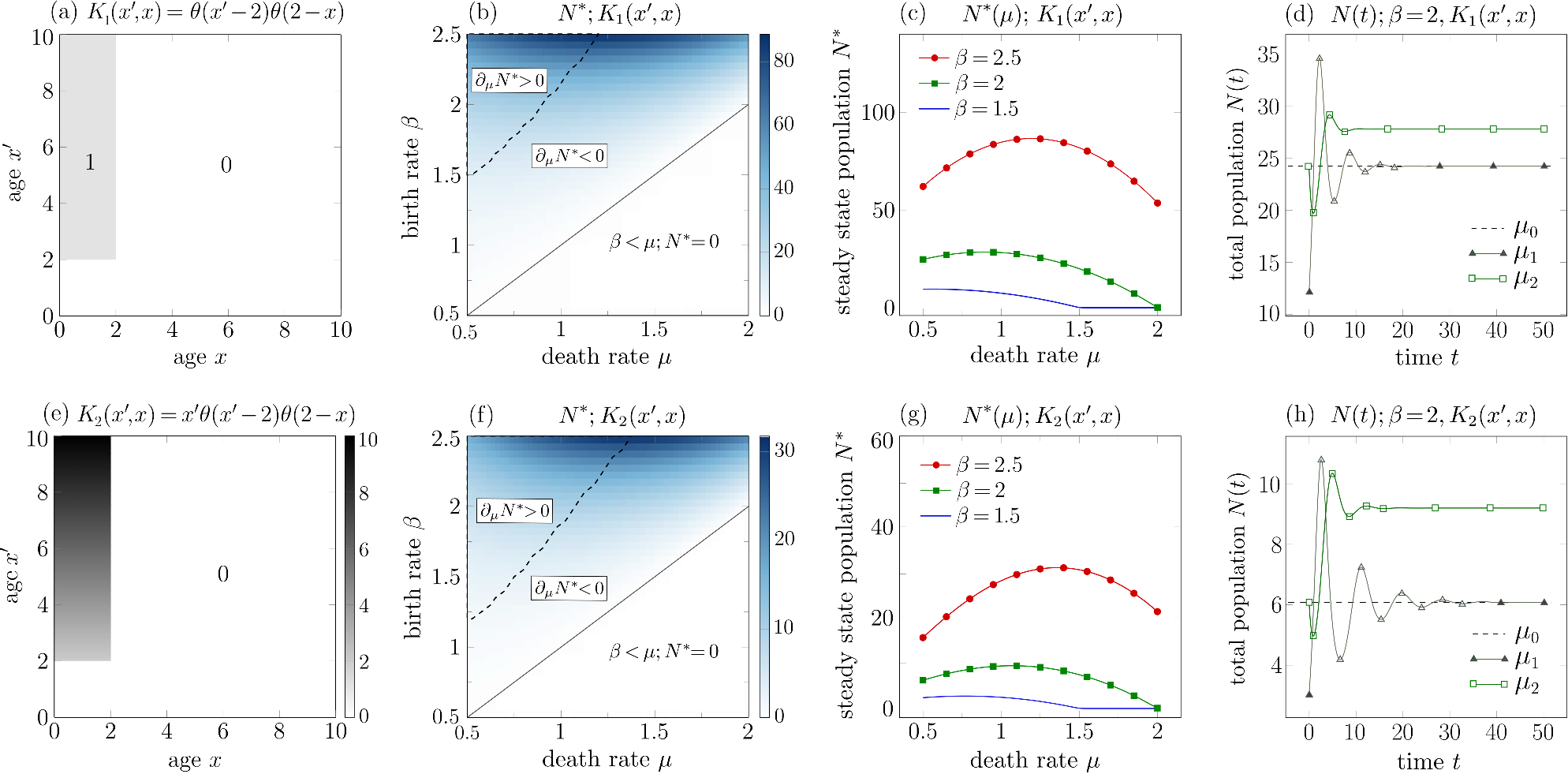}
\caption{\small {(a) Heatmap of the dimensionless predation
    interaction $K_1(x', x)=\theta(x'-2)\theta(2-x)$.  (b) Heatmap of
    the total steady-state population $N^{*}$ as a function of
    constant $\beta$ and $\mu$.  A nontrivial stable fixed point
    arises only for $\beta > \mu$.  The region of no overcompensation,
    where $\partial_{\mu}N^{*} <0$, is indicated while the parameters
    that admit steady-state overcompensation, where
    $\partial_{\mu}N^{*}> 0$ (not indicated), occur in the upper-left
    corner. The dashed curve delineates the phase boundary on which
    $\partial_{\mu}N^{*}= 0$. (c) $N^{*}$ plotted as a function of
    $\mu$ for fixed values of $\beta = 2.5,2,1.5$. (d) Plots of $N(t)$
    for $\beta = 2$ and death rate sequences $\mu_{1}(t)$ and
    $\mu_{2}(t)$ (Eq.~\eqref{mu_3}).  For $\mu_{1}(t)$, damped
    oscillations yield transient overcompensation, while $\mu_{2}(t)$
    results in a permanent, steady-state overcompensation, in addition
    to damped oscillations. (e-h) The corresponding results for $\beta
    = 2$ and the predation/cannibalization interaction $K_{2}(x',x) =
    x' K_{1}(x',x)$.}}
\label{over_popu_fig}
\end{figure*}

Since analytically finding all conditions under which
Eq.~\eqref{LVmodel} or Eq.~\eqref{steady_state} exhibits
overcompensation is difficult, we shall carry out numerical
experiments to show how overcompensation arises for some simple forms
of $K(x',x)$ after instantaneous changes in $\beta$ and $\mu$ from one
constant value to another. In general, we find that a cannibalism
interaction $K$ that decreases with $x$ and increases with $x'$ is
more likely to exhibit larger overcompensation.
%
%
We examine two simple forms of $K$: $K_{1}(x', x)=k
\theta(x'-X)\theta(X-x)$ and $K_{2}(x', x)= kx' K_{1}(x',x) =
k^{2}x'\theta(x'-X)\theta(X-x)$, both of which satisfy
Eq.~\eqref{eq:compact_condition}. Since $k$ is a rate, we can measure
$\beta$ and $\mu$ in units of $k$ and time $t$ and ages $x$ in units
of $1/k$.  In such units, we set $k=1$ without loss of generality and
the dimensionless interactions take the forms
\begin{equation}
\begin{aligned}
  K_{1}(x', x) & \equiv  \theta(x'-X)\theta(X-x),\\
  K_{2}(x', x) & \equiv  x' K_{1}(x',x)\\
 \: &  = x'\theta(x'-X)\theta(X-x).
\end{aligned}
\end{equation}
For concreteness, we choose $X=2$ and plot heatmaps of the
dimensionless predation kernels $K_{1}$ and $K_{2}$ in
Figs.~\ref{over_popu_fig}(a) and (e).  Subsequent results derived from
using these interactions are displayed across each row.

The analysis of steady-state overcompensation boils down to
investigating how the solution $n^{*}$ obeying
Eq.~\eqref{steady_state}, and in particular, how the total population
$N^{*}=\int_0^{\infty}n^{*}(x)\dd{x}$ changes with $\beta$ and $\mu$.
Figs.~\ref{over_popu_fig}(b) and (f) plot $N^{*}$ and the solution to
Eq.~\eqref{steady_state} using $K_{1}$ and $K_{2}$, respectively.
We see that for both types of interactions, regimes in which
$\partial_{\mu}N^{*}>0$ arise, signaling permanent
overcompensation. In Figs.~\ref{over_popu_fig}(b) and (f), the
parameter regime exhibiting overcompensation is shown in the upper
left when birth rates are large and death rates small.  The dashed
curves Figs.~\ref{over_popu_fig}(b) and (f) mark the ``phase
boundary'' of overcompensation at which $\partial_{\mu}N^{*} = 0$.
For larger $\mu$ and smaller $\beta$, $\partial_{\mu}N^{*} < 0$, and
there is no overcompensation. Note that when $\beta < \mu$, the only
stable state is $n^{*}(x), N^{*}=0$. Figs.~\ref{over_popu_fig}(c) and
(g) show the corresponding curves $N^{*}(\mu)$ for fixed values of
$\beta$, quantitatively illustrating the different magnitudes of
steady-state overcompensation through different values of the slope
$\partial_{\mu}N^{*}$. These results, along with the interactions
shown to preclude long-lasting overcompensation, indicate that
permanent overcompensation in our model requires cannibalization of
the young by the old and a $K(x',x)$ that increases in $x'$ and
decreases in $x$.

To interrogate the dynamics of the population following perturbations
to the death rate, we now start the system at its steady state
corresponding to specific values $\beta_{0},\mu_{0}$ and consider how
the population $N(t)$ evolves after applying these two different
perturbations:

  \begin{equation}
    \mu_1(t)  = \mu_{0} + (\log 2) \delta(t), \quad
\mu_2(t) = \mu_{0} + \theta(t)\Delta\mu 
\label{mu_3}.
    \end{equation}
%
To be specific, we take $\beta_{0}=2$, $\mu_{0}=1/2$, and $\Delta\mu =
1/2$. The death rate function $\mu_{1}(t)$ includes a delta function
at $t=0$, which corresponds to an instantaneous removal of half the
population from the steady state associated with $\beta_{0},\mu_{0}$ and
the corresponding interaction $K$. A finite volume discretization
\citep{eymard2000finite} with $\Delta x=0.02, x_{\max}=10, \Delta
t=0.002$ was used to numerically solve Eq.~\eqref{LVmodel} to find
$n(x,t)$, which is then used to construct $N(t)
=\int_{0}^{\infty}n(x,t)\dd x$. Figs.~\ref{over_popu_fig}(d) and (h)
show damped oscillations in $N(t)$ associated with $K_{1}$ and $K_{2}
= x' K_{1}$, respectively. Although $\mu_{1}(t)$ immediately returns
to the value $\mu_{1}(t>0) = \mu_{0}$, and $N(t\to\infty) \to N^{*}$,
at shorter times, $N(t)$ oscillates and can exceed $N^{*}$ at
intermediate times. Thus, \textit{transient overcompensation} can
arise even though the population returns to the same value set by
$\beta_{0}, \mu_{0}$. If $\mu_{2}(t)$ is used, the death rate jumps
from $\mu_{0}$ to $\mu_{0}+\Delta \mu$ at $t=0$, leading ultimately to
a higher steady-state population. For $\mu_{2}$, in addition to a
higher steady-state population, initial oscillations can lead to even
higher transient populations.

Motivated by these results showing that $N^{*}$ can increase upon
increasing $\mu$ for fixed values of $\beta$, we provide in Appendix
\ref{app:misc} additional examples of mechanisms whereby a
steady-state overcompensation can arise. First, we consider
overcompensation in a model where resources are scare and predation
can provide nourishment required for reproduction.  This effect can be
modeled by a predation-enhanced fecundity of the form $\beta(x) =
\beta_0 + (\tfrac{1}{4})\int_0^{\infty}\!\!K(x, x')n(x',
t)\dd{x'}$. This model is shown to preserve overcompensation
associated with increases in $\mu$, as detailed in Appendix
\ref{app:birthK}.

We also show in Appendix \ref{app:harvesting} that age-dependent
harvesting can lead to overcompensation. Harvesting or culling of the
population that is modeled via an additional removal term
\begin{equation}
\begin{aligned}
  \frac{\partial n(x,t)}{\partial t} + \frac{\partial n(x,t)}{\partial x}
  = & -\bigg[\mu(x, t) + \int_{0}^{\infty}\!\! K(x', x)n(x', t)
    \dd{y}\bigg] n(x,t) \\
  \: & - h(n; x, t),\\
n(0, t) &  =  \int_0^{\infty}\!\!\beta(x)n(x, t)\dd{x}, 
\end{aligned}
\label{LVmodel_harvest}
\end{equation}
where $h(n; x, t)$ represents the rate of harvesting that may depend
nonlinearly on the structured population.  Increases in a realistic
harvesting function $h(n; x, t)$ are shown to lead to permanent
increases in the total population. Finally, we also prove in Appendix
\ref{birth_OC} that for an interaction that satisfies
Eq.~\eqref{eq:compact_condition}, increasing a constant $\beta$ will
always lead to an increase in $N^{*}$; however, for asymmetric
predation kernels that can be negative (a {young-eat-old}
interaction), overcompensation in response to increased birth rates,
where $\partial_{\beta}N^{*} <0$ for fixed values of $\mu$, can arise.
 
\subsection{Undamped oscillations}
\label{example2}
The instantaneous changes in the death rate given by Eqs.~\eqref{mu_3}
and the interactions $K_{1}$ and $K_{2}$ give rise to damped
oscillations that eventually settle back to their corresponding unique
values $N^{*}$.  However, oscillations may be undamped and lead to
\textit{periodic overcompensation} when the fixed point loses
stability and bifurcates to a stable limit cycle. Such oscillations
have been observed, for example, in European green crabs populations
\citep{grosholz2021stage}. Although the source of such oscillations
may be difficult to disentangle from the effects of seasonality, they
have been modeled in different contexts using a single-compartment
\textit{discrete-time} population model \citep{boyce1999seasonal}.
Overcompensation has also been described in consumer-resource models,
as cycles of rising and falling populations
\citep{pachepsky2008between}, as in the classical predator-prey model.

Here, we use a simple, realistic old-eat-young cannibalization rate

\begin{equation}
    K_{3}(x', x) = (x'-x)\theta(x'-2)\theta(2-x)
\label{cannibalism_rate_oscillation}
\end{equation}
in Eq.~\eqref{LVmodel} and assume constant birth rate $\beta$ and
death rate $\mu$.  Upon using a finite volume discretization with
$\Delta x=0.01$ and initial condition $n(x, 0) = e^{-2x}/2$, we
numerically solve Eq.~\eqref{LVmodel} in $t \in [0,T], \, \Delta t =
0.002$ to investigate whether the total population $N(t)$ oscillates.
\begin{figure*}[htb]
\centering
\includegraphics[width=6.6in]{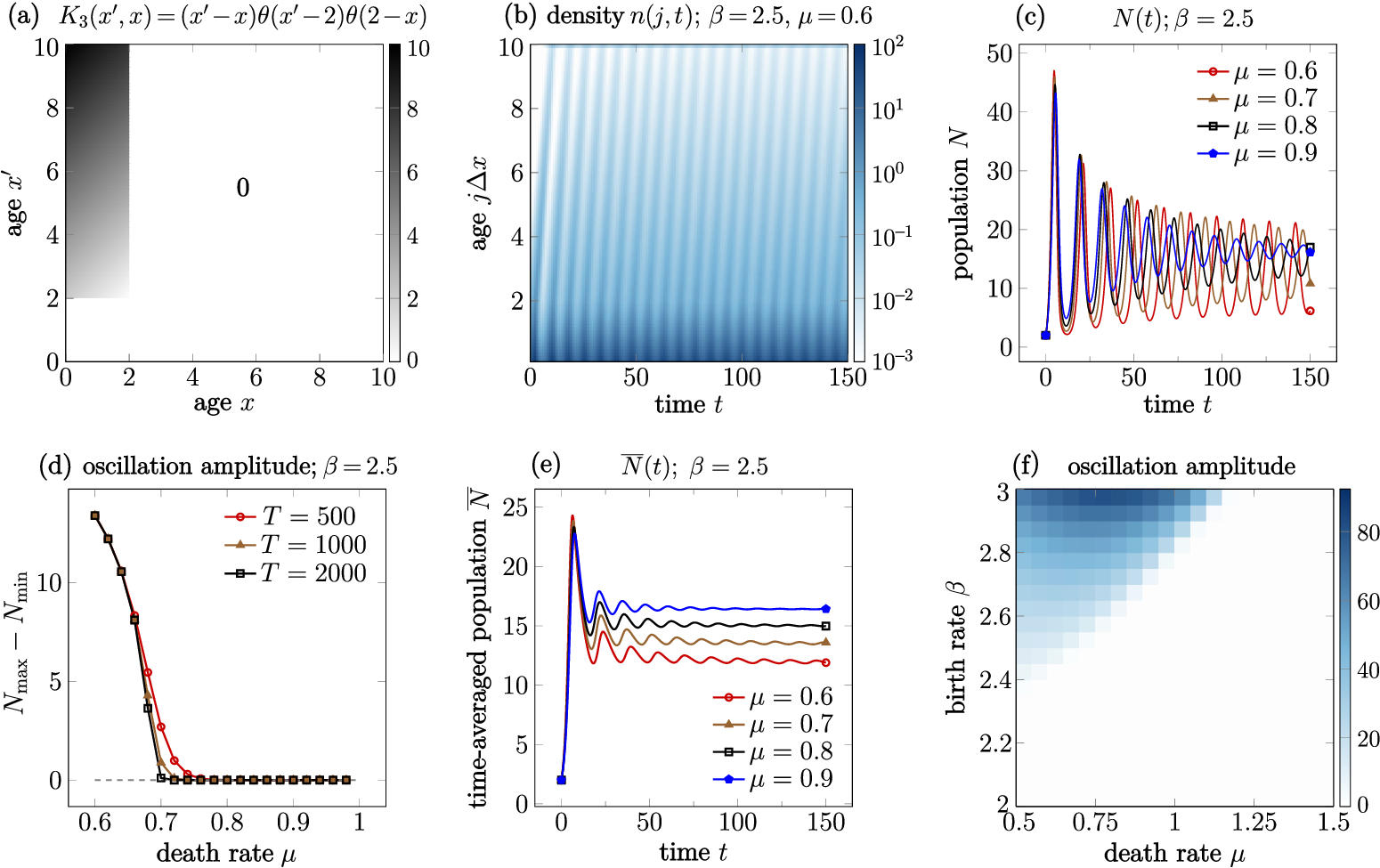}
\caption{\small {(a) Heatmap of the interaction kernel $K_{3}(x',x) =
    (x'-x)\theta(x'-2) \theta(2-x)$
    (Eq.~\eqref{cannibalism_rate_oscillation}).  (b) The population
    density computed using Eq.~\eqref{cannibalism_rate_oscillation},
    $\beta=2.5$, and $\mu=0.6$, and approximated as $n(j,t)\equiv
    (\Delta x)^{-1}\int_{j\Delta x}^{(j+1)\Delta x} n(y,t)\dd y$ with
    $\Delta x = 0.02$ displays persistent periodic oscillations.  (c)
    The total population $N(t)=\int_0^{\infty}n(x,t)\dd{x}$ also
    exhibits oscillations that persist(damp out) for small(large) values
    of $\mu$.  (d) The long-time oscillation amplitude $\max_{T - 20
      \leq t \leq T}N(t) - \min_{T - 20 \leq t \leq T}N(t)$ near
    $T=500$, $T=1000$, and $T=2000$, respectively, plotted as a
    function of $\mu$ ($\beta = 2.5$). As $T$ is increased, the
    transition to oscillating states as $\mu$ is decreased becomes
    sharper These numerical results suggest that the transition at
    $\mu \sim 0.7$ has jump discontinuity in the first-order
    derivative. (e) Time-averaged populations $\overline{N}(t)\equiv
    \tfrac{1}{t} \int_{0}^{t} N(s)\dd{s}$ for $\beta=2.5, \mu=0.6,
    0.7, 0.8, 0.9$ that more clearly reveal the mean values of $N(t\to
    \infty)$. (f) Oscillation amplitude in $\beta$-$\mu$ space.  As
    $\beta$ increases from $2$ to $3$ and $\mu$ decreases from $1.5$
    to $0.5$, undamped oscillations arise. Here, the oscillation
    amplitudes are measured by $\max\limits_{480<t\leq500}N(t) -
    \min\limits_{480<t\leq500}N(t)$. In the regime plotted, we find
    that undamped oscillations arise for $\beta \gtrsim
    1.87+0.93\mu$.}}
\label{oscillatory}
\end{figure*}
Fig.~\ref{oscillatory}(a) shows the heatmap of the interaction kernel
$K_{3}(x',x) = (x'-x)\theta(x'-2) \theta(2-x)$, while
Fig.~\ref{oscillatory}(b) shows a heatmap of an oscillating structured
population density $n(x,t)$ approximated by its local mean value
$n(j,t) = (\Delta x)^{-1}\int_{j\Delta x}^{(j+1)\Delta x}n(y,t)\dd
y$. These oscillations lead to an oscillating total population $N(t)$,
as shown in Fig.~\ref{oscillatory}(c). Oscillations damp out when
$\mu$ is large, but persist for smaller values of $\mu$. The long-time
amplitudes of oscillation shown in Fig.~\ref{oscillatory}(d) indicate
a sharp decrease as $\mu$ is increased. To better resolve the
long-time average values of $N(t)$, we define its function average
$\overline{N}(t) \equiv \tfrac{1}{t}\int_{0}^{t} N(s) \dd s$ and plot
them in Fig.~\ref{oscillatory}(e). Besides oscillations that can lead
to temporary overcompensation, increasing $\mu$ in the regime studied
also led to increased averaged values of $N(t)$, and in particular,
when oscillations are damped out at larger $\mu$, permanent
overcompensation can arise where the steady values $N^{*}$ increase
with $\mu$. Thus, as $\mu$ increases, periodic overcompensation
transitions to steady-state overcompensation. The phase diagram
separating regimes of transient and permanent oscillations is shown in
Fig.~\ref{oscillatory}(f). As $\beta$ increases and $\mu$ decreases,
the dynamics transition from a monotonically converging one (to
steady-state value $N^{*}$) to a periodically oscillating one, with a
finite oscillation magnitude that arises when $\beta$ exceeds a
critical value $\beta_{*}\approx 1.87+0.93\mu$.

\subsection{Reduction of structured population PIDE to ODE systems}
\label{ODEmodel}

We have provided some numerical examples which explicitly show various
types of overcompensation in response to variations in constant $\mu,
\beta$.  However, our model can also be approximated via
coarse-graining and discretization and formulated in terms of a system
of coupled nonlinear ODEs.  Systems of ODEs are typically used to
describe multispecies models in which previous studies have found
overcompensation. Multistage models in which, \textit{e.g.}, adult or
later-stage insect feed on eggs or early-stage individuals
\citep{MANICA_2003,OHBA_2006} can also be directly modeled by our
discrete stage discretized ODEs.

Since the analysis of the general nonlinear PIDE model
Eq.~\eqref{LVmodel} or the steady-state integral-differential equation
Eq.~\eqref{steady_state} is difficult and uniqueness only under
Eq.~\eqref{eq:compact_condition} and a few specific proofs of cases
that preclude overcompensation could be found (see Appendix
\ref{NOOC}), related analyses of the ODE system can be more easily
performed \citep{de2008simplifying,sorenson2021intra} if parameters
and variables are considered to be piecewise constants. In addition to
providing mathematical insight into the approximate, lower dimensional
ODE system, the simplest numerical implementation of a finite volume
method for the PIDE model Eq.~\eqref{LVmodel} is conceptually similar
to piecewise constant discretization in the age variable.

Here, we formally discretize our PIDE model and explore whether the
resulting ODE models exhibit the analogous behaviors of the full PIDE
model discussed above. We discretize the space of ages $[0,\infty)$
  into $L+1$ bins: $[x_i,x_{i+1}),~ i=0,...,L$ where $x_i= i\Delta x$
    if $i \leq L$, and $x_{L+1}=\infty$. Let the population on the
    $i^{\textrm{th}}$ bin $[x_i,x_{i+1})$ be denoted $n_{i}(t) =
      \int_{x_i}^{x_{i+1}} n(y,t) \dd y, \,\, (x_{L+1}\to \infty)$. By
      integrating Eq.~\eqref{LVmodel} over increments, each $n_i$
      obeys
\begin{equation}
\begin{aligned}
\frac{\dd n_i}{\dd t} = &\,  n(x_i, t) - n(x_{i+1}, t)
- \int_{x_i}^{x_{i+1}}\mu(x, t)n(x, t)\dd{x} \\
\: & \quad  - \int_{x_i}^{x_{i+1}} \!\!\!\dd x \int_x^{\infty} \!\! 
\dd x'\, n(x', t)K(x', x)n(x, t), \\
n(0, t) = & \sum_{i=0}^{L}\int_{x_i}^{x_{i+1}}\!\! \beta(x,t) n(x, t)\dd{x}.
\end{aligned}
\label{piecePDE}
\end{equation}
This reduction technique is similar to that used in
\citep{diekmann2020finite} to reduce infinite dimensional PDE models
for structured populations into a finite-dimensional set of ODEs.

We now take the coefficients $\beta, \mu$, and $K$ to be piecewise
constant in each compartment, \textit{i.e.},
\begin{equation}
\begin{aligned}
\beta(x,t) & = \beta_i(t), \qquad x\in[x_i, x_{i+1}), \\
\mu(x, t) & = \mu_i (t), \qquad x\in[x_i, x_{i+1}),\\
K(x', x) & = K_{j, i}(t), \quad\,  x'\in[x_j, x_{j+1}),\, x\in[x_i, x_{i+1}), 
\end{aligned}
\end{equation}
where $\beta(x,t)=\beta_i(t),\,\, x\in[x_i, x_{i+1})$ if $\beta$
  is independent of cannibalism.  Eq.~\eqref{piecePDE} then simplifies
  to
%

\begin{equation}
\begin{aligned}
  \frac{\dd n_0}{\dd t}  = & - \mu_0(t)n_0( t) - n_0(t)\sum_{j=0}^{L}\! K_{j,0} n_j(t)
  +\!\sum_{j=1}^{L}\beta_{j}(t)n_{j}(t) - \frac{n_{0}(t)}{\Delta{x}}, \\
  \frac{\dd n_i}{\dd t}  = & - \mu_{i}(t)n_{i}(t) - n_i(t)\sum_{j=i}^{L}\!K_{j,i}n_j(t)
  -\frac{n_i(t) -n_{i-1}(t)}{\Delta{x}}, \\
%
\frac{\dd n_{L}}{\dd t} = & - \mu_{L}(t)n_{L}(t) - n_{L}(t)
K_{L,L} n_{L}(t)+\frac{n_{L-1}(t)}{\Delta{x}}.
\end{aligned}
\label{pieceODE}
\end{equation}
$K_{i,i}$ represents the within-compartment competition introduced due
to the discretization. In the following, we will assume that
$K_{i,i}=0$ and that $\beta_0=0$. Note that the ODE system
Eq.~\eqref{pieceODE} is also the discretized finite volume method we
used to numerically solve the original PIDE Eq.~\eqref{LVmodel}.  We
are particularly interested in whether the simplified ODE model
Eq.~\eqref{pieceODE} with time-independent coefficients gives rise to
the rich dynamics observed in the original PIDE model, especially as
$L$ is varied.  Our main results are: (i) the ODE system
Eq.~\eqref{pieceODE} has at most one positive steady state, (ii) the
two-compartment ODE model (setting $L=1$ in Eq.~\eqref{pieceODE}) has
a unique positive steady state and the steady-state populations
$n_{0}$ and $n_{1}$ never increase with the death rate. This result
differs from that of the two-stage ODE model in
\citep{sorenson2021intra} because a different type of interaction was
invoked, (iii) the three-compartment ODE model ($L=2$) exhibits a
unique, positive, stable steady state and can exhibit overcompensation
of increased death rates, and (iv) higher-$L$ ODE systems can exhibit
long-term oscillations in addition to permanent overcompensation as
the positive steady state destabilizes.  The proofs for these results
are detailed in Appendix \ref{DISCRETE}.

\section{Summary and Conclusions}
\label{conclusion}

In this paper, we use a linear kinetic population model to formally
derive a bilinear, age-structured PIDE that incorporates a continuum
of intraspecies cannibalistic interactions.  Distinct from previous
models that typically assume complicated interactions within
multistage/multispecies populations or rely on complex
consumer-resource interactions, we demonstrate mathematically that our
single-species, bilinear interaction model, structured simply
according to age, can give rise to a variety of dynamical behavior.

Although similar PIDEs have been previously proposed
\citep{Hastings_Costantino1987,Hastings1987,TAKASHINA_FIKSEN,TAKASHINA}
and undamped population oscillations found, overcompensation in
response to increased death were not treated. We used realistic forms
of predation to show that this model can exhibit permanent,
steady-state overcompensation of the total population in response to
permanent increases in death rate. General forms of predation kernels
$K(x',x)$ that preclude steady-state overcompensation were enumerated
showing that gradients in both $x'$ and $x$ are necessary for static
overcompensation (when $\beta$ and $\mu$ are constants).
Specifically, our analysis suggests that $K(x',x)$ that increases in
$x'$ and decreases in $x$ are more likely to exhibit steady-state
overcompensation. Using predation kernels $K_{1}(x',x)$ and
$K_{2}(x',x)$, Eq.~\eqref{LVmodel} was solved numerically using a
finite volume method to show the emergence of steady-state
overcompensation. Our PIDE model can also be numerically solved using
recently developed spectral methods that adaptively decompose the
solution $n(x,t)$ into spatial basis functions with time-dependent
coefficients.  These methods are detailed in
\citep{xia2021efficient,xia2021frequency, chou2022adaptive} and are
quite efficient at handling unbounded domains.

Our analyses also allowed us to quantitatively distinguish transient
overcompensation from steady-state overcompensation. Dynamic, or
transient overcompensation was defined in terms of oscillations in the
total population that also arose under predation kernels $K_{1}$ and
$K_{2}$ and abrupt changes in the values of $\mu$ and $\beta$ (see
Fig.~\ref{over_popu_fig}). These cases exhibited damped oscillations
in the total population that transiently exceeded their expected
steady-state values. At long times, the total populations converged to
steady values uniquely associated with their permanent values of
$\mu$. For $\mu$ that has permanently increased, steady-state
overcompensation is not universal but arises only under certain values
of $\beta$ and $\mu$. However, for values of $\beta$ and $\mu$ under
which steady-state overcompensation \textit{does not} arise (for
$K=K_{1}, K_{2}$), transient overcompensation may nonetheless arise
following jumps in $\mu(t)$.

Using certain forms of $K$ (see Fig.~\ref{over_popu_fig}), dynamic or
transient overcompensation was observed in terms of oscillations in
the total population that eventually damps to steady values that could
be lower or higher (steady-state overcompensation) following increases
in $\mu(x)$. However, similar to predator-prey models that can exhibit
periodic oscillations, we also found that an interaction such as
$K_{3}(x',x) = (x'-x)\theta(x'-2) \theta(2-x)$ leads to undamped
oscillations in total population for certain values of $\beta$ and
$\mu$. We found numerically that permanent oscillations emerge in a
way suggestive of a Hopf bifurcation as $\mu$ is decreased.  It would
be interesting to develop analytic results for how stability is gained
or lost as $\mu$ is tuned.

Besides formal proofs that certain simple predation interactions rule
out permanent overcompensation, and numerical exploration of specific
cases that exhibit dynamical (damped and undamped oscillations) and
steady-state overcompensation, a rigorous analysis of our nonlinear
structured population PIDE model remains elusive. However,
simplification via coarse-graining and discretizing the age variable
allowed the PIDE to be cast as a system of approximating ODEs for
piecewise constant parameter functions $\beta(x), \mu(x)$, and
$K(x',x)$.

Revutskaya \textit{et al.} \citep{discrete_time_hydra} has shown that
a discrete-time, two-sex, three-compartment model exhibits
multistability and overcompensation under harvesting. However,
overcompensation of increased death does not require the presence of
multistability. Under certain conditions, the ODEs derived from our
original cannibalistic-interaction PIDEs showed at most one positive
steady state, implying that permanent overcompensation of increases in
the death rate in our model cannot be due to transitions from one
steady state to another. In our formulation, steady-state
overcompensation and permanent oscillations are also recapitulated in
ODE systems of at least three and four dimensions, respectively. These
results may provide insight into mathematical strategies for analyzing
our PIDE model under age-dependent birth and death rates.

Our mathematical framework suggests a number of possible future
avenues of investigation. For example, since chaotic behavior has been
shown to arise in a three-dimensional, two-species predator-prey ODE
model \citep{wikan2021compensatory}, an intriguing question is how
chaotic solutions might arise in our single-species continuously
structured model Eq.~\eqref{LVmodel}. Continuously structured PIDE
models can also be combined within multicomponent/multispecies models
where even richer behavior might arise.  For example, multicompartment
aging models with \textit{symmetric} age-age interactions have been
shown to give rise to waves in opinion dynamics
\citep{RADICALIZATION}. How overcompensation or oscillatory behavior
of the total population when it is structured according to and evolves
in size (Eq.~\eqref{deterministic_growth}) rather than age is also
worthwhile modeling. Finally, in analogy with spatial predator-prey
models \citep{cosner1999effects,cantrell2001dynamics}, including
age-dependent spatial diffusion within our continuum structured PIDE
model may lead to intriguing behavior such as transport-mediated local
and global overcompensation.


%

\section*{CRediT authorship contribution statement}
Mingtao Xia: Conceptualization, Formal analysis, Investigation,
Methodology, Validation, Visualization, Writing – original draft,
Writing – review \& editing. Xiangting Li: Conceptualization, Formal
analysis, Investigation, Methodology, Validation, Visualization,
Writing – original draft, Writing – review \& editing. Tom Chou:
Conceptualization, Formal analysis, Investigation, Project
Administration, Supervision, Methodology, Validation, Visualization,
Writing – original draft, Writing – review \& editing.

\section*{Declaration of competing interest}
The authors declare that they have no known competing financial
interests or personal relationships that could have appeared to
influence the work reported in this paper.

\section*{Data availability}
No data was used for the research described in the article.


\bibliographystyle{unsrt} 

\bibliography{bibliography_overcompensation}


\section*{Mathematical Appendices}

\appendices


\section{Interactions that preclude permanent overcompensation}
\label{NOOC}

Here, we consider a few explicit forms for the cannibalism rate
$K(x',x)$
%
%
that are analytically tractable. We prove that these simple
interaction terms preclude overcompensation of increases in death
rate.

\subsection{Self-inhibition $K(x',x)=k(x)\delta(x'-x)$}
\label{i}

We first show that if cannibalization occurs within individuals of the
same structured variable (age in this case), \textit{i.e.}, $K(x', x)=
k(x)\delta(x'-x), k(x)>0$, no overcompensation can occur, even for
age-dependent birth and death rates $\beta(x)$ and $\mu(x)$.  The
steady-state Eq.~\eqref{steady_state} becomes a Riccati equation with
a specific boundary condition,

\begin{equation}
\begin{aligned}
{\dd  n^{*}(x) \over \dd x} & = -\mu(x) n^*(x) - k(x)(n^*(x))^2, \\
n^*(0) & = \int_0^{\infty}\!\!\beta(x)n^*(x)\dd{x}.
\end{aligned}
\label{case_one}
\end{equation}
After defining $q^*(x) \coloneqq k(x) n^*(x)$, Eq.~\eqref{case_one}
simplifies to
\begin{equation}
\begin{aligned}
\frac{\dd q^{*}(x)}{\dd x} & = -\hat{\mu}(x) q^*(x) - [q^*(x)]^2,\,\,\,
\hat{\mu}(x) \coloneqq \mu(x)-\frac{k'(x)}{k(x)}, \\
q^*(0) & = \!\int_0^{\infty}\! \frac{k(0)}{k(x)}\beta(x) q^*(x)\dd{x}.
\end{aligned}
\label{case_two}
\end{equation}
Substituting $q^*(x) = u'(x)/u(x)$ into Eq.~\eqref{case_two}, we
obtain the linear ODE $\frac{\dd^{2} u(x)}{\dd x^{2}} +
\hat{\mu}(x)\frac{\dd u(x)}{\dd x} = 0$ which admits the general
solution
\begin{equation}
    u(x) \propto \big(1+C \int_0^x \! e^{-\int_{0}^{z}\hat{\mu}(y) \dd{y}}\dd{z}\big),
\end{equation}
where $C$ is an integration constant and it is assumed that
$\hat{\mu}$ such that $\int_{0}^{z}\hat{\mu}(y) \dd{y}$ has a finite
lower bound and that $e^{-\int_{0}^{z}\hat{\mu}(y) \dd{y}}$ is
integrable. The steady-state population density $n^*(x)$ is then
reconstructed as

\begin{equation}
    n^*(x) = \frac{1}{k(x)}\frac{e^{-\int_{0}^{x}\hat{\mu}(x')\dd x'}}
{\tfrac{1}{C} +
  \int_0^x \!e^{-\int_0^z\hat{\mu}(y)
    \dd{y}}\dd{z}}, \quad C = k(0)n^{*}(0).
      \label{case_one_sol}
\end{equation}
Substituting Eq.~\eqref{case_one_sol} into Eq.~\eqref{case_one}, 
we find the constraint on $C = k(0)n^*(0)$

\begin{equation}
  1 = \int_0^{\infty}\frac{k(0)}{k(x)}  \frac{\beta(x)e^{-\int_0^{x}\hat{\mu}(x')\dd{x'}}}
  {1+C\int_0^x \!e^{-\int_0^z\hat{\mu}(y)\dd{y}}\dd{z}}\, \dd{x}.
\end{equation}

Suppose we have two different death rates $\mu_1(x) \geq \mu_2(x)$
(and thus $\hat{\mu}_1(x) \geq \hat{\mu}_2(x)$) with their
corresponding steady-state solutions $n_1^*(x), n_2^*(x)$ defined by
their integration constants $C(\mu_1), C(\mu_2)$.  We first show that
$C(\mu_1)>C(\mu_2)$. Define


\begin{equation}
  F_{\mu}(C) = \int_0^{\infty}
  \frac{k(0)}{k(x)} \frac{\beta(x)e^{-\int_0^x\hat{\mu}(x') \dd{x'}}}
{1+C\int_0^x \! e^{-\int_0^z\hat{\mu}(y)\dd{y}}\dd{z}}\,\dd{x}, 
\end{equation}
which is a decreasing function of $C$ when $C>0$. 
Next, note that 

\begin{equation}
  \frac{e^{-\int_0^x \hat{\mu}_{1}(x')\dd{x'}}}{1+C\int_0^x
    \!e^{-\int_0^z\hat{\mu}_{1}(y)\dd{y}}\dd{z}} 
  \leq \frac{e^{-\int_{0}^{x}\hat{\mu}_{2}(x')\dd{x'}}}
       {1+C\int_0^x\! e^{-\int_{0}^{z}\hat{\mu}_{2}(y)\dd{y}}\dd{z}}
\end{equation}
if $\mu_1(x)\geq \mu_2(x)$. Thus, $F_{\mu_1}(C(\mu_1)) = 1$ and
$\mu_1(x) \geq \mu_2(x)$ imply $F_{\mu_2}(C(\mu_1)) >
1$. Together with the constraint $F_{\mu_2}(C(\mu_2))=1$ and
monotonicity of $F_{\mu_2}$, $F_{\mu_2}(C(\mu_1)) > 1$ implies
$C(\mu_{2}) > C(\mu_{1})$; in other words,
$n_2^*(0)>n_1^*(0)$. Furthermore, we have for all  $x \geq 0$

\begin{equation}
  \begin{aligned}
  n_1^*(x) & =  \frac{1}{k(x)}\frac{e^{-\int_0^x \hat{\mu}_{1}(x') \dd{x'}}}{\tfrac{1}{C(\mu_1)}+\int_0^x 
    \! e^{-\int_0^z\hat{\mu}_{1}(y) \dd{y}}\dd{z}} \\
  \: & \leq
\frac{1}{k(x)}\frac{e^{-\int_0^x \hat{\mu}_{2}(x')\dd{x'}}}{\tfrac{1}{C(\mu_2)}
  +\int_0^x \! e^{-\int_0^z\hat{\mu}_{2}(y)\dd{y}}\dd{z}} = n_2^*(x).
\end{aligned}
\end{equation}
Thus, the total populations $N_1^*$ and $N_2^*$ satisfy
$N_1^*=\int_0^{\infty}\!\!n_1^*(x)\dd{x}\leq
\int_0^{\infty}\!\!n_2^*(x)\dd{x}=N_2^*$.  We conclude that no
overcompensation will be observed under an interaction of the form
$K(x'-x)=k(x)\delta(x'-x)$.


\subsection{$x$-independent cannibalism rate $K=K(x')$}
\label{ii}

We also show that an $x$-independent predation interaction (predators do not
prefer prey of any age), $K(x',x) = K(x')$, precludes permanent
overcompensation.  In this proof however, we must assume
age-independent birth and death $\beta(x)=\beta, \mu(x)=\mu$.
%
%
For $K(x',x) = K(x')$, the solution to Eq.~\eqref{steady_state}
satisfies
\begin{equation}
\begin{aligned}
n^{*}(x) & = n^{*}(0)e^{-(\mu+K^*)x}, \quad K^*\!\coloneqq \!\int_0^{\infty}\!\!K(x')n^{*}(x')\dd{x'}, \\
n^{*}(0) & = \beta N^*\! = \beta \!\int_{0}^{\infty}\!\!n^{*}(x)\dd x.
    \end{aligned}
\label{nii}
\end{equation}
%
%
When $\mu \geq \beta$, no positive $K^{*}$ in Eq.~\eqref{nii} can
satisfy $N^{*} = \int_{0}^{\infty}n^{*}(x)\dd x$ and no positive
solution exists. Numerical integration of the full time-dependent
model in Eq.~\eqref{LVmodel} shows that the only steady state is $n^*
\equiv 0$. When $\mu < \beta$, the solution to Eq.~\eqref{nii} is
satisfied by $K^* = \beta - \mu$ which leads to $n^{*}(x) = \beta N^*
e^{-\beta x}$. Upon using $n^{*}(x) = \beta N^* e^{-\beta x}$ in the
expression $K^* = \beta - \mu = \int_{0}^{\infty} \!  K(x') n^*(x')\dd
x' = \beta N^{*} \int_{0}^{\infty}K(x')e^{-\beta x'}\dd x'$, we find

\begin{equation}
    N^{*}= 
\frac{1-\tfrac{\mu}{\beta}}{\int_0^{\infty}\! K(x')e^{-\beta{x'}}\dd{x'}},
\end{equation}
which strictly decreases with $\mu$. Thus, a predation kernel that is
independent of prey age $x$ cannot exhibit steady-state
overcompensation.


\subsection{$x'$-independent cannibalism rate $K(x',x)=K(x)$}
\label{iii2}

For a predation/cannibalization rate of the form $K(x',x) = K(x)$, the
steady-state Eq.~\eqref{steady_state} becomes
\begin{equation}
\begin{aligned}
\displaystyle \frac{\dd n^{*}(x)}{\dd x} 
&= -\bigg[\mu + K(x)\!\int_0^{\infty}\!\!n^{*}(x')\dd{x'}\bigg]n^{*}(x), \quad
\\ n^*(0)  &= \beta N^*.
\end{aligned}
\label{niii2}
\end{equation}
We now prove that if $\mu, \beta$ are constants, then no permanent
overcompensation will occur.  Equation~\eqref{niii2} is solved by
$n^{*}(x) = n^{*}(0)e^{-\mu x - N^{*}\tilde{K}(x)}$, where
$\tilde{K}(x) \equiv \int_{0}^{x}\!K(y)\dd y$.  Upon integrating the
solution and using the boundary condition $n^{*}(0) = \beta N^{*}$,
eliminating $n^{*}(0)$, and using the definition
$N^{*}=\int_{0}^{\infty}\! n(x)\dd x$, we find an implicit solution
for $N^{*}$:

\begin{equation}
1 = \beta \!\int_{0}^{\infty}\!\! e^{-\mu x - N^{*}\tilde{K}(x)}\dd x  \equiv F(\mu, N^{*}).
\label{F2=1}
\end{equation}
Eq.~\eqref{F2=1} is the specific form of Eq.~\eqref{eq:Euler-Lotka} to
be derived under a general condition later. To see how $N^{*}$ varies
with $\mu$, we apply the implicit function theorem to obtain

%

\begin{equation}
\displaystyle \frac{\delta N^{*}}{\delta \mu} = - \frac{(\partial
  F/\partial \mu)}{(\partial F/\partial N^{*})} =
-\frac{\int_{0}^{\infty}\!xe^{-\mu x - N^{*}\tilde{K}(x)}\dd x}
{\int_{0}^{\infty}\! \tilde{K}(y) e^{-\mu y - N^{*}\tilde{K}(y)}\dd y}.
\end{equation}
For $\mu,N^{*} > 0$, the RHS above is negative. Thus,
$\partial N^{*}/\partial \mu < 0$ and steady-state overcompensation cannot
arise. This result implies that an interaction kernel $K(x',x)$ that
varies only in $x$ is insufficient for steady-state overcompensation
and that variation in $x'$ is necessary. This result, along with that
in section \ref{ii}, suggests that predation kernels $K(x',x)$ that
vary in both $x'$ and $x$ are required for steady-state
overcompensation, at least for age-independent $\beta$ and $\mu$.


\section{Simple interactions that can exhibit steady-state overcompensation}
\label{YESOC}

Here, we consider some solvable examples of predation kernels
$K(x',x)$ that \textit{can} exhibit steady-state overcompensation.

\subsection{Point source and sink}
\label{sec:point_source_sink}
Consider the predation kernel defined as \( K(x',x) = k \delta(x'-b)
\delta(x-a) \), and $\beta$, $\mu$ assumed to be constant.
The steady state solution of such a system satisfies:
\begin{equation} \label{eq:steady_state}
\frac{\dd n^*(x)}{\dd x} = -\Big[\mu + k n^*(b) \delta(x-a)\Big] n^*(x).
\end{equation}

For such a system, the steady-state expression for $n^*(x)$ can be
written as:
\begin{equation} \label{eq:n_t}
n^*(x) = n^*(0) e^{-\mu x - k n^*(b) \theta(x-a)}.
\end{equation}
Assuming $b > a$, we deduce from the boundary condition that
\begin{equation} \label{eq:constriant}
1 = \int_0^{\infty} \beta e^{-\mu x - k n^*(b) \theta(x-a)} \dd x.
\end{equation}
Since $n^{*}(b)>0$, $(1-e^{-kn^{*}(b)}) < 1$, and 
Eq.~\eqref{eq:constriant} can be satisfied if and only if
\begin{equation} \label{eq:positive_solution}
e^{-\mu a} > 1 - \frac{\mu}{\beta}.
\end{equation}
Under this condition, $n^*(b)$ can be determined as 
\begin{equation}
  \label{eq:n_b_behavior}
n^*(b) = -\frac{1}{k}\ln\Big(1-\big(1-\tfrac{\mu}{\beta}\big)e^{\mu a}\Big),
\end{equation}
which, when used in conjunction with Eq.~\eqref{eq:n_t} determines 
$n^*(0)$, leading to the final expression for the steady state total
population
\begin{equation} \label{eq:N_star_expression}
N^* = -\frac{e^{\mu b}}{\beta k}\frac{\ln
  \Big(1-\big(1-\tfrac{\mu}{\beta}\big)e^{\mu
    a}\Big)}{\Big(1-\big(1-\tfrac{\mu}{\beta}\big)e^{\mu a}\Big)}.
\end{equation}
One can see that the magnitude of interaction strength $k$ does not change the 
sign of $\partial N^*/\partial \mu$. In the limit $a \rightarrow 0^{+}$,
Eq.~\eqref{eq:N_star_expression} reduces to
\begin{equation} 
\label{eq:N_star_expression_a_0}
N^*(a \rightarrow 0^{+}) = \frac{e^{\mu b}}{\mu k} 
\ln\left(\tfrac{\beta}{\mu}\right).
\end{equation}
Note that in this limit, the constraint
Eq.~\eqref{eq:positive_solution} is always satisfied, although we
still require $\beta > \mu$ to ensure that $N^*$ is positive.  Taking
the derivative of Eq.~\eqref{eq:N_star_expression_a_0} with respect to
$\mu$, we find
\begin{equation} \label{eq:dN_star_dmu}
\frac{\partial N^*(a \rightarrow 0^{+})}{\partial \mu}
= \frac{e^{b \mu}}{\mu^2 k} \left[(b \mu -1) \ln\Big(\tfrac{\beta}{\mu}\Big)-1\right].
\end{equation}
In other words, $\partial N^*/\partial \mu$ is positive, with
overcompensation, if and only if
\begin{equation} \label{eq:overcompensation_condition}
  (\mu b -1) \ln\big(\tfrac{\beta}{\mu}\big) > 1.
\end{equation}
Since $\mu \in (0, \beta)$, the lowest value of $b$ that can still
yield $\partial N^*/\partial \mu > 0$ (overcompensation) for some
values of $\beta, \mu$ is
\begin{equation} 
\label{eq:overcompensation_condition_b}
  b_{\rm c}= \inf_{\mu \in (0,\, \beta)}\left(\frac{1}{\mu} 
+ \frac{1}{\mu \ln \frac{\beta}{\mu}}\right) = 
  \frac{e^{-\frac{1}{2}+\frac{\sqrt{5}}{2}}}{\beta}
+\frac{e^{-\frac{1}{2}+\frac{\sqrt{5}}{2}}}{\beta\left(-\frac{1}{2}+\frac{\sqrt{5}}{2}\right)}.
\end{equation}
Intuitively, large birth rate $\beta$ and large interaction distance
$b>b_{\rm c}$ are both favorable for overcompensation.

\subsection{Predation on newborns/eggs}

We now delve into a specific version of our older-predating-on-younger
model, where the predation kernel $K$ is given by

\begin{equation}
K(x',x) = \lim_{\varepsilon \to 0^{+}}k\delta(x-\varepsilon) \theta(x'-b),
\label{eq:predation_kernel}
\end{equation}
where $\theta$ is the Heaviside function.  In this scenario, the
steady state solution $n^*(x)$ is solved by
\begin{equation}
n^*(x) = n^*(0)e^{-k N_b} e^{- \mu x},
\label{eq:steady_state_adult_eat_egg}
\end{equation}
where
\begin{equation}
N_b\!\coloneqq\!\int_b^{\infty}\!\!n^*(x)\mathrm{d} x \equiv \frac{n^*(b)}{\mu}.
\label{eq:Nb_def}
\end{equation}
The birth (boundary) condition can then be expressed as

\begin{equation}
1 = \tfrac{\beta}{\mu}e^{-kn^{*}(b)/\mu}.
\label{eq:birth_condition}
\end{equation}
Given the constraint $\beta>\mu$, the steady state population at $x=b$
becomes $n^*(b) = (\mu/k) \ln\Big(\beta/\mu\Big)$. Using this result
in Eq.~\eqref{eq:steady_state_adult_eat_egg} allows us to determine
$n^*(0)$ and total population $N^*$

\begin{equation}
N^* = \frac{e^{\mu b}}{k} \ln\Big(\tfrac{\beta}{\mu}\Big).
\label{eq:total_population}
\end{equation}
To demonstrate there exists an interval of $\mu$ for which
overcompensation arises provided sufficiently large $b$, we examine
when the condition

\begin{equation}
\frac{\partial N^*}{\partial \mu} = \frac{\left(\mu b \ln
  \left(\frac{\beta}{\mu}\right)-1\right) e^{\mu b}}{\mu k} > 0
\label{eq:overcompensation_condition_adult_eat_egg}
\end{equation}
holds. The threshold of $b$ above which overcompensation is possible
is thus

\begin{equation}
b_{\rm c}= \inf_{\mu \in (0,\, \beta)} \frac{1}{\mu \ln\Big(\tfrac{\beta}{\mu}\Big)} = \frac{e}{\beta}.
\label{eq:threshold_b}
\end{equation}


\section{Uniqueness of the positive steady state of Eq.~\eqref{LVmodel}}
\label{UNIQUE}

If the distributed interaction $K(x',x)$ satisfies
Eq.~\eqref{eq:compact_condition}, we shall prove uniqueness of a
positive steady state under the assumption that the set $\{x:\exists
x'>x>0, K(x',x)>0\}\cap \{x:\beta(x)>0, x>0\}$ has positive measure.
We shall prove the following two statements. First, we assume two
steady states, $m^{*}(x)$ and $n^{*}(x)$, and demonstrate that if
$m^{*}(X)=n^{*}(X)$ at some age $X$, then $m^{*}(x)$ and $n^{*}(x)$
are precisely the same steady state everywhere.  Second, without loss
of generality, if $n^{*}(X)>m^{*}(X)$, we will demonstrate that
$n^{*}(x)>m^{*}(x)\,\, \forall x \geq 0$. This dominance relation
conflicts with the well-known Euler-Lotka equation, thereby
demonstrating the uniqueness of the steady-state solution.  These
results indicate that although overcompensation can arise from
transition to an alternative steady state upon increases in death or
harvesting \citep{discrete_time_hydra}, multistability is not a
necessary condition for overcompensation of increased death.
  
To show
\begin{equation}
	m^*(X) = n^*(X) \Rightarrow m^*(x) = n^*(x), \,\,\forall x \geq X,
\label{x>X}
\end{equation}
first note that since $K(x',x>X) = 0$, the interaction terms in
Eq.~\eqref{steady_state} for both $m(x)$ and $n(x)$ vanish for $x>X$
and thus are linear first-order equations with identical decay rates
$\mu(x)$ and coincident ``initial conditions'' $m^*(X) =
n^*(X)$. Thus, the solutions for $x>X$ are identical.

What remains is to show that $m^*(X)=n^*(X) \Rightarrow
m^*(x)=n^*(x),\, \forall x \geq 0$.  To simplify notation, we set $\xi
= X-x$, $\xi' = X-x'$ and define $f(\xi)\equiv n(X-x)$,
$f^{*}(\xi)\equiv n^{*}(X-x)$, transforming the steady-state problem
Eq.~\eqref{steady_state} into a general integral-differential equation
with given initial data (using Eq.~\eqref{x>X})

\begin{equation}
\begin{aligned}
{\dd f(0< \xi\leq X)\over \dd \xi} & \displaystyle  = \left[ \mu(\xi)+\!\!\int_{-\infty}^{\xi} 
\!\!K(\xi',\xi)f(\xi')d \xi' \right]f(\xi), \\
f(\xi\leq 0 ) & = f^{*}(\xi),
\end{aligned}
\label{backward_ODE}
\end{equation}
where we have reparameterized $\mu(x)$ such that $\mu(\xi) =
\mu(x=X-\xi)$ and $K(x',x)$ such that $K(\xi',\xi)= K(x'=X-\xi',
x=X-\xi)$.  The goal is to march the steady-state uniqueness from $\xi
< 0$ ($x>X$) up to $\xi = X$ ($x=0$). Let us assume that uniqueness of
$f(\xi)$ has been demonstrated up to $\xi_{0}$, \textit{i.e.}, that
$f(\xi) = f^{*}(\xi)\equiv n^{*}(X-\xi)$ has been uniquely determined
in $(-\infty, \xi_{0}]$. Breaking up the integral term, we write
\begin{equation}
\begin{aligned}
\displaystyle  {\dd f(\xi> \xi_{0})\over \dd \xi} = & 
\bigg[\mu(\xi) +\!\! \int_{-\infty}^{\xi_{0}} 
\!\! K(\xi',\xi)f(\xi')\mathrm{d} \xi' \\
\: & \quad\qquad\qquad \displaystyle 
+ \!\!\int_{\xi_{0}}^{\xi} \!\!K(\xi',\xi)f(\xi')\mathrm{d} \xi' \bigg] 
f(\xi),\\
f(\xi\leq \xi_{0}) = & f^{*}(\xi).
\end{aligned}
\label{backward_ODE_SPLIT}
\end{equation}
We can march $\xi_{0}$ forward from 0 and consider a small region
$(\xi_{0}, \xi_{0}+\varepsilon)$ successively.  At each stage, since
$f(\xi) = f^{*}(\xi_{0} \leq \xi <\xi_{0}+\varepsilon)$ has been
uniquely determined, we can combine
$\mu(\xi)+\int_{-\infty}^{\xi_{0}}K(\xi',\xi)f(\xi')\dd \xi' \to
\mu(\xi)$.  If we start at $\xi_0=0$, it suffices to show that the
solution to
\begin{equation}
\begin{aligned}
\frac{\dd f(\xi>0)}{\dd \xi} 
= & \left[\mu(\xi) + \int_{0}^{\xi} \!K(\xi',\xi)f(\xi')\mathrm{d} \xi' \right] 
f(\xi),\\
f(0) = & f^{*}(0)
\end{aligned}
\label{original_2nd}
\end{equation}
%
is unique in a small domain of $(0,\varepsilon)$.

Suppose that $\mu(\xi)$ is (locally) bounded by $\hat{\mu}$,
$K(\xi',\xi)$ is bounded by $\hat{K}$, and $g$ is the local solution
to the associated differential equation

\begin{equation}
\begin{aligned}
 \frac{\dd g(\xi>0)}{\dd \xi} 
= & \left[\hat{\mu} + \hat{K}\!\!\int_{0}^{\xi}\!\!g(\xi')\mathrm{d} \xi' \right]
g(\xi), \\
g(0) = & f^{*}(0).  
\end{aligned}
\end{equation}
The integral $G(\xi)\equiv \int_{0}^{\xi} g(\xi')\mathrm{d} \xi'$ then
obeys the standard-form second-order ODE

\begin{equation}
\begin{aligned}
\frac{\dd^{2} G(\xi>0)}{\dd \xi^{2}} = &
\left[\hat{\mu} + \hat{K}G(\xi)\right] \frac{\dd G(\xi)}{\dd \xi},\\
\frac{\dd G(\xi)}{\dd \xi}\bigg|_{\xi=0}\!\!\! = &\, g(0) \equiv f^{*}(0),\,\,\,  G(0) = 0. 
\end{aligned}
\label{2ndode}
\end{equation}
The solution to Eq.~\eqref{2ndode} in the region $(0, \varepsilon)$ is
unique and for any solution $f$ of Eq.~\eqref{original_2nd}, $0< f
\leq g$.

Now suppose that $f_1$ and $f_2$ are two solutions in the neighborhood of $0$
that solve Eq.~\eqref{original_2nd}. We have
\begin{equation}
\begin{aligned}
\hspace{-2mm}f_i(\xi)\!=\!f^{*}(\xi)+ & \!\!\int_{0}^{\xi}\!\bigg(\mu(\xi')+\!\!\int_{0}^{\xi'}\!\!\!\!K(\xi'', \xi') f_i(\xi'')\mathrm{d}\xi''\bigg)f_i(\xi')d\xi'\label{f_SOLN}
\end{aligned}
\end{equation}
for $i=1, 2$. Note that
\begin{align}
\begin{split}
& \bigg| f_1(\xi')\!\int_{0}^{\xi'}\!\!K(\xi'',\xi') f_1(\xi'') \dd \xi'' 
 \\[-4pt]&\qquad\qquad\quad - f_2(\xi')\!\int_{0}^{\xi'}\!\!K(\xi'',\xi') f_2(\xi'') \dd \xi'' \bigg| \\
& \hspace{0.7cm} = \bigg|f_1(\xi')\!\int_{0}^{\xi'}\!\!K(\xi'',\xi')
\Big(f_1(\xi'')-f_{2}(\xi'')\Big)\dd \xi'' 
\\[-4pt] &\qquad\qquad\,\, + \Big(f_1(\xi')-f_{2}(\xi')\Big)\!\int_{0}^{\xi'}\!\!K(\xi'',\xi')f_{2}(\xi'') \dd \xi'' \bigg| \\
& \hspace{0.8cm}  \leq 2\hat{K} \xi' \sup_{\xi\in(0,\varepsilon)}\!\!g(\xi)
\sup_{\xi\in (0,\varepsilon)}\!\big|f_1(\xi)-f_2(\xi)\big|.
\end{split}
\end{align}
Then, using Eq.~\eqref{f_SOLN}, we conclude that
\begin{equation}
\begin{aligned}
\sup_{\xi\in [0,0 + \varepsilon]}\!\!\big|f_1(\xi)-f_2(\xi) \big| \leq &
\Big\{ \varepsilon \hat{\mu} + \varepsilon^2 
\hat{K}\!\sup_{\xi\in (0,\varepsilon)}\!g(\xi) \Big\}\!\!\sup_{\xi\in [0,0+ \varepsilon]}\!\!\!\big|f_1(\xi)-f_2(\xi)\big|.
\end{aligned}
\end{equation}
Since $\varepsilon$ can be chosen sufficiently small such that $\big\{
\varepsilon \hat{\mu}\!+\!\varepsilon^2 \hat{K} \sup_{\xi\in
  (0,\varepsilon)}\!g(\xi) \big\}<1$, we conclude $\sup_{\xi\in [0,0+
    \varepsilon]}\big| f_1(\xi)-f_2(\xi) \big|=0$, proving the
solution to Eq.~\eqref{original_2nd} is unique in a neighborhood of
$0$. Under the assumption that the solution to
Eq.~\eqref{original_2nd} exists, we can replace the point $\xi=0$ with
$\xi\in(0, K)$ and conclude that the solution is unique in a small
neighborhood of $\xi$.  Therefore, the solution is globally unique in
$(0, K)$, and the proof of the first statement is completed.

Next, we prove the second statement by showing that the case
$n^*(X)>m^*(X)$ cannot not hold by first claiming that
\begin{equation}
	n^*(X) > m^*(X) \Rightarrow n^*(x) > m^*(x), \,\,\forall x \geq 0.
\end{equation}
We easily observe that the statement is true for $x \geq X$. Suppose
for some $x_0$, $n^*(x_0) \leq m^*(x_0)$.  Then let $x^*=\sup_{x\geq
  0}\{ x: n^*(x) \leq m^*(x) \}$. By continuity of $m^*$ and $n^*$, we
note that $x^*< X$.

Within the interval $(x^*,X)$, $n^*(x)>m^*(x)$. Let $\xi^*=X-x^*$, and
consider the functions $m^*$ and $n^*$ written as functions of
$\xi$. The difference of Eq.~\eqref{backward_ODE}
satisfied by $m^*$ and $n^*$ becomes
\begin{equation}
\frac{\dd }{\dd \xi} \left(n^*(\xi)-m^*(\xi) \right) 
\geq {\mu(\xi)} \left(n^*(\xi)-m^*(\xi) \right), \,\,\,\forall \xi \in (0,\xi^*).
\label{eq:special_inequality}
\end{equation}
By integrating both sides of Eq.~\eqref{eq:special_inequality} from
$0$ to $\xi^*$, we conclude that $n^*(\xi^*)-m^*(\xi^*)>0$, which
demonstrates the dominance relation $n^*(x) > m^*(x),\, \forall x \geq
0$.

We can now exploit the equilibrium form of the Euler-Lotka equation
\citep{Keyfitz1997sep,Sharpe1911apr}. Let
$\tilde{\mu}(x)=\mu(x)+\int_{x}^{\infty} K(x',x)n^*(x')dx'$ be the
effective death rate and $s(x)=\exp \left(-\int_0^{x}\tilde{\mu}(x')
\dd x'\right)$ be the survival probability of any individual up to age
$x$.  The overall steady-state rate of new births defined by

\begin{equation}
n(0,t)\equiv B(t) =\!\int_{0}^{\infty}\!\!\beta(x)n(x,t)\dd x
\label{B}
\end{equation}
can be formally written in terms of $s(x)$ and the form of the method
of characteristics solution $n(x<t,t) = n(0,t-x)s(x)=
B(t-x)s(x)$. Using this form of $n(x<t,t)$ in the integrand in
Eq.~\eqref{B}, we find in the $t\to \infty$ limit $B(t) =
\int_{0}^{\infty}\!\!\beta(x) B(t-x)s(x)\dd x$.  Since in the $t \to
\infty$ steady state limit all quantities are independent of time,
$B(t) = B$ and

\begin{equation}
1 = \int_{0}^{\infty}\!\! \beta(x)s(x)\dd x\equiv\!\int_{0}^{\infty}\!\! \beta(x)
e^{-\int_{0}^{x}\tilde{\mu}(x')\dd x'}\dd x,
\label{eq:Euler-Lotka}
\end{equation} 
which means that at steady state, the population $n^{*}(x)$ and
effective death rate $\tilde{\mu}(x')$ settles to value such that the
effective reproductive number $R_{0}\equiv
\int_{0}^{\infty}\!\!\beta(x)s(x)\dd x = 1$.

Equation~\eqref{eq:Euler-Lotka} must be satisfied at steady state but
allows us to compare different $\tilde{\mu}$s associated with
different steady state solutions.  For different steady states $m^*$
and $n^*$ such that $m^*>n^*$ at all ages, the effective death rates
satisfy $\tilde{\mu}^{(m)} \geq \tilde{\mu}^{(n)}$. Since $\beta$
remains the same, the survival probability satisfies $s^{(m)} \leq
s^{(n)}$, where the inequality should hold on a positive measure
interval. Because $\{x:\exists x'>x>0, K(x',x)>0\}\cap \{x:\beta(x)>0,
x>0\}$ has positive measure, we will conclude
\begin{equation}
1 =\!\!\int_{0}^{\infty}\!\!\!\beta(x)e^{-\int_0^x
  \tilde{\mu}^{(m)}(x')\mathrm{d}x'} \mathrm{d}x <\!\!\int_{0}^{\infty}\!\!\!\beta(x) e^{-\int_0^x
  \tilde{\mu}^{(n)}(x')\mathrm{d}x'}\!\mathrm{d}x\!=\!1.
\end{equation}
This contradiction shows that if $K$ is continuous and compactly
supported and if $\{x:\exists x'>x>0, K(x',x)>0\}\cap \{x:\beta(x)>0,
x>0\}$ has positive measure, then Eq.~\eqref{LVmodel} admits at most
one positive steady state.


\section{Existence of a positive steady state of Eq.~\eqref{LVmodel}}
\label{EXISTENCE}

If a predation kernel satisfies Eq.~\eqref{eq:compact_condition}, we
can also obtain the criterion for the existence of a positive steady
state, which is equivalent to finding a positive solution $n^*(x)$ to
Eqs.~\eqref{steady_state} under certain additional assumptions.

In Appendix~\ref{UNIQUE}, we showed that any solution $n^*(x)$ to
Eqs.~\eqref{steady_state} must satisfy Eq.~\eqref{original_2nd} with the
transformed coordinate $\xi = X -x$ and $f(\xi)= n^*(X-x)$. For the
existence arguments, we first investigate the condition under which
Eq.~\eqref{original_2nd} has a positive solution. Formally, we pick
the initial condition $n_X \coloneqq f(0) > 0$ as the parameter of
interest, and consider the domain of $n_X$ such that the solution to
Eq.~\eqref{original_2nd} exists up to $\xi = X$.  Define $f_{-1}(\xi)
\equiv 0$ and $f_{n+1}(\xi)$ as solutions to the ODE

\begin{equation}
\begin{aligned}
  \frac{1}{f_{n+1}(\xi)}\frac{\mathrm{d} f_{n+1}(\xi)}{\mathrm{d} \xi}
  & = \mu(\xi) +\!\int_{0}^{\xi}\!\!K(\xi',\xi) f_{n}(\xi')\mathrm{d}\xi',\,\, 
\forall \xi \in (0,X),\\ 
f_{n+1}(0) &= n_X,
\end{aligned}
\label{iterate_ODE}
\end{equation}
where $n \geq -1$. In particular, $f_0(\xi)=n_X e^{\int_0^{\xi}
  \mu(\xi')\mathrm{d} \xi'} > 0 = f_{-1}(\xi)$ for all $\xi \in
(0,X)$.  For each $n$, an iterative argument shows that $f_n$ is
bounded, continuous, and nonnegative on $[0,X]$. In addition, $f_0
(\xi) > f_{-1}(\xi), \forall \xi \in (0,X)$ implies that
$\{f_n(\xi)\}$ is a monotonically increasing sequence in both $\xi$
and $n$. Therefore, $f(\xi):= \lim_{n \rightarrow \infty} f_n(\xi) \in
{(0, \infty]}$ exists and satisfies Eq.~\eqref{original_2nd} up to the
  moment of blowup $\xi^* \equiv \sup\left\{\xi \in (0,X): f(\xi) <
  \infty \right\}$, thanks to the monotone convergence theorem.

We also observe that $f(\xi)$ depends monotonically on the initial
value $n_X$. For sufficiently regular $\mu$ and $K$, we also assume
that $f(\xi)$ depends continuously on $n_X$. Define the upper limit
for $n_X$ by $n_X^* \equiv \sup \left\{ n_X: \xi^*(n_X) > X \right\}$
with the convention that $\sup \emptyset = 0$, and we find an open
domain $(0, n_X^*)$ of $n_X$ such that $f(\xi) < \infty$ for all
$\xi\in [0,X]$ with the initial value $n_X$. In particular, the
continuity assumption implies $\lim_{n_X \rightarrow n_{X^{-}}^{*}}
f(X)= \infty$. The marginal case $n_X^* = \infty$ is covered by this
equation because $f(X) \geq f(0)= n_X \rightarrow \infty$.

Now, we recover $n^*(x)$ from $f(\xi)$ and denote $n^*$ by $n^*_{n_X}$
to emphasize the dependence on $n_X$. For the sake of simplicity, we
assume the upper limit $n_X^*= + \infty$ in the following
discussion. This can be achieved by imposing proper restrictions on
$\mu$ and $K$, such that the existence of the solution to
Eq.~\eqref{2ndode} on the interval $(0,X)$ is guaranteed.

We proved that there is a unique solution $n^*_{n_X}(x)$ to the first
equation in Eqs.~\eqref{steady_state} when $n^*_{n_X}(X) = n_X$
provided that $\mu$ and $K$ are bounded on $[0,X]$, there exist
positive constants $\mu_0, K_0>0$ such that $\mu\geq\mu_0, K\geq K_0$,
and $K$ vanishes for $x>X$.  We shall show that the solution to
Eqs.~\eqref{steady_state} exists if (i) in the cannibalism-free
environment ($K=0$), the expected number of offspring that
an individual will give birth to is larger than 1

\begin{equation}
 \int_0^{\infty}\!\!\beta(x)e^{-\int_0^x \mu(x')\mathrm{d} x'} \mathrm{d}x > 1,
\label{expected_offspring}
\end{equation}
and (ii) given any $\frac{\dd G(\xi)}{\dd \xi}\big|_{\xi=0}\coloneqq
n_X$, the second order ODE Eq.~\eqref{2ndode} can be solved up to
$\xi=X$, and that $\frac{\dd G(\xi)}{\dd \xi},\, \xi\in(0, X)$ depends
continuously on the initial $\frac{\dd G(\xi)}{\dd \xi}|_{\xi=0}$.

The existence of the solution to Eqs.~\eqref{steady_state} is then
converted to finding a proper $n_X$ such that $n^*_{n_X}(0)$ is
suitable for the second equation in Eqs.~\eqref{steady_state},
\textit{i.e.}, the boundary condition representing the newborn cells.
So we need to show that $\frac{\dd G(\xi)}{\dd \xi}\big|_{\xi=X} =
n^*_{n_X}(0)$ satisfies the boundary condition in
Eqs.~\eqref{steady_state}. Note that as long as $n_X > 0$,
$n^*_{n_X}(x)>0, ~\forall x \geq 0$. Let $\tilde \mu$ denote the
effective death rate $\tilde{\mu}(x) = \mu(x)+\int_x^{\infty} K(x',
x)n^*_{n_X}(x)\dd{x}$, then the second equation in
Eqs.~\eqref{steady_state} is equivalent to

\begin{equation}
n^*_{n_X}(0) \left\{ 1 - \int_0^{\infty}\!\!\beta(x) e^{-\int_0^x
  \tilde{\mu}_{n_X}(x')\mathrm{d} x'}\mathrm{d} x \right\} = 0.
\end{equation}
We define the Euler-Lotka functional as

\begin{equation}
\textrm{EL}\big[n^*_{n_X}\big] 
= \int_0^{\infty}\!\! \beta(x)e^{-\int_0^x \tilde{\mu}_{n_X}(x')\mathrm{d}x'}
\mathrm{d} x,
\label{functional}
\end{equation}
Then, the second equation in Eq.~\eqref{steady_state} is equivalent to
the famous Euler-Lotka equation $\textrm{EL}\big[n^*_{n_X}\big] = 1$ for
positive solutions $n^*_{n_X}$. We've shown that $n_X < n_X'
\Rightarrow \tilde{\mu}_{n_X}(x) \leq \tilde{\mu}_{n_X'}(x), ~ \forall
x \geq 0$. Therefore, the function $n_X \mapsto
\textrm{EL}\big[n_{n_X}^*\big]$ is monotonically decreasing. Because
$n^{*}_{n_X}(x),\, x\leq X$ depends continuously on $n^*_{n_X}$, the
functional Eq.~\eqref{functional} also depends continuously on
$n^*_{n_X}$. Consequently, we conclude that the existence of the
positive steady state is equivalent to

\begin{equation}
\begin{aligned}
\lim_{n_X \rightarrow \infty} \int_0^{\infty}\!\beta(x) e^{-\int_0^x 
\tilde{\mu}_{n_X}(x') \mathrm{d}x'} \mathrm{d}x  & < 1; \\
\lim_{n_X \rightarrow 0+}\int_0^{\infty}\!\beta(x)e^{-\int_0^x 
\tilde{\mu}_{n_X}(x') \mathrm{d} x'}\mathrm{d}x & > 1.
\end{aligned}
\label{2a}
\end{equation}

When $n_X=0$, $n^*_{n_X} (x) \equiv 0$. Furthermore, for $x < X$,
$n^*(x) \geq n^*(X)$, which implies $\lim_{n_X \rightarrow
  \infty}n^*_{n_X}(x)= \infty, x<X$.  Since we have assumed that both
$\mu$ and $K$ have positive lower bounds on their support and that the
solution $n^*_{n_X}$ is continuously dependent on the initial
condition $n_X$, we could conclude that

\begin{equation}
\begin{aligned}
\lim_{n_X \rightarrow 0+} \tilde{\mu}_{n_X}(x) = &\mu(x) +\!\!\int_x^{\infty}\!\!\!K(x',x) n^*_0(x')\mathrm{d} x'\! = \mu(x);\\
\lim_{n_X \rightarrow \infty} \tilde \mu_{n_X} (x) = &\mu(x)+\!\!\int_x^{\infty}\!\!\!K(x',x) n_{\infty}^*(x')\mathrm{d} x\! = \begin{cases}
\mu(x) & \!\!x\geq X \\
\infty & \!\!\textrm{otherwise.}
\end{cases}
\end{aligned}
\end{equation}
The first equation in Eq.~\eqref{2a} is satisfied as 

\begin{equation}
\lim_{n_X \rightarrow \infty} \int_0^{\infty}\!\beta(x) e^{-\int_0^x
  \tilde{\mu}_{n_X}(x')\mathrm{d} x'}\mathrm{d} x = 0
\end{equation}
because $\lim_{n_X \rightarrow \infty} \tilde \mu_{n_X} (x) =\infty,
x\in(0, X)$. Furthermore, the second equation in Eq.~\eqref{2a} is
satisfied by the assumption Eq.~\eqref{expected_offspring}. Therefore,
there must exist an $n_X$ such that ${\rm EL}\big[n_{n_X}^*\big]=1$, and thus the
corresponding $n_{n_X}(x)$ satisfies the two equations in
Eqs.~\eqref{steady_state}.


\section{Additional examples of overcompensation}
\label{app:misc}

\subsection{Cannibalism-related birth rate}
\label{app:birthK}

\begin{figure*}[htb]
\centering
\includegraphics[width=7in]{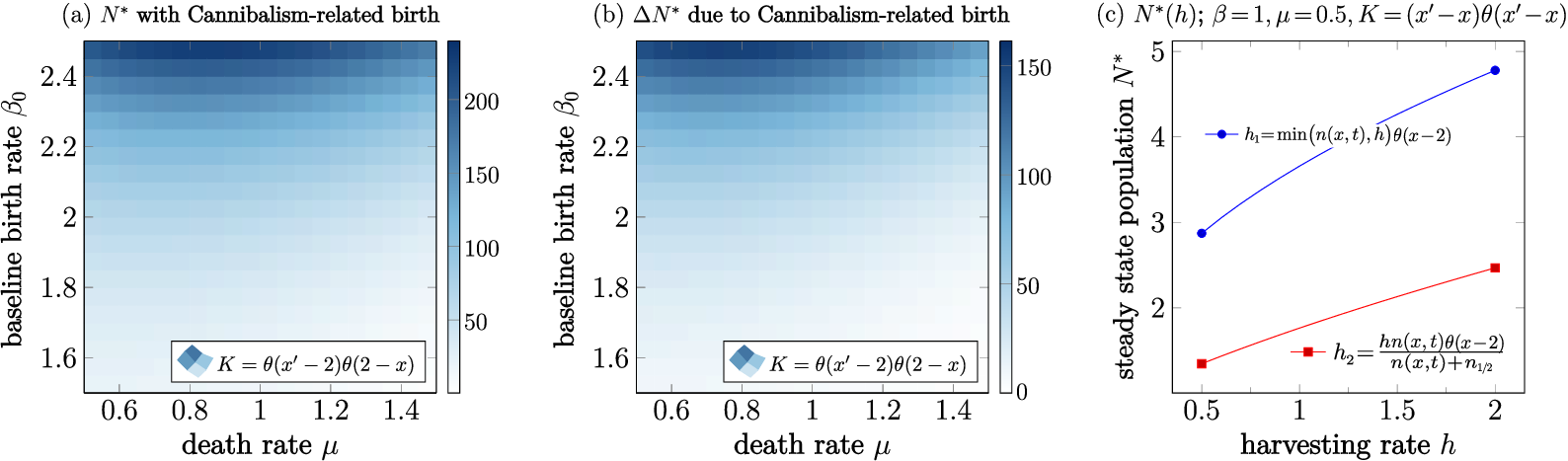}
\caption{\small (a) The steady-state total population $N^*(\beta(x,
  \phi[n;x]))$ that displays overcompensation with a constant death
  rate for the cannibalism-dependent birth rate
  Eq.~\eqref{cannibalism_related}, where cannibalism has a positive
  effect on the birth rate. (b) The difference in the steady-state
  population $N^*(\beta(x, \phi[n;x])) - N^*(\beta_0)$, where
  $N^*(\beta(x, t))$ is the steady-state total population with a
  constant birth rate $\beta\coloneqq\beta_0$. Because $\beta(x,
  \phi[n;x])>\beta_0$ for $n^*$, the difference is always positive.
  Furthermore, for some fixed $\beta_0$, the difference $N^*(\beta(x,
  \phi[n;x])) - N^*(\beta_0)$ also ``overcompensates" by first
  increase then decrease with $\mu\coloneqq\mu_0$. (c) The
  steady-state total population for $h_1, h_2$ in Eq.~\eqref{h_def}
  for the harvesting model Eq.~\eqref{LVmodel_harvest}.
  Overcompensation is observed with increasing harvesting rates.
  Furthermore, since $h_1>h_2$ if $h$ is the same, for a fixed $h$,
  the total population under the harvesting rate $h_1$ is greater than
  that under $h_2$.\label{harvest}}
\end{figure*}

In the main discussion, we assumed that cannibalism only modifies the
death rate. Here, we provide a numerical example in which preying on
juveniles can increase birth rates. This limit may arise when food is
not abundant and cannibalism provides nourishment for reproduction.
In this case the birth rate can be a function of the amount of
cannibalism measured by $\int_0^{\infty}K(x, x')n(x', t)\dd{x'}\equiv
\phi[n;x]$, which is a functional of $n(x)$.  We will assume that 
$\beta(x, \phi[n;x])$ takes the form 

\begin{equation}
\begin{aligned}
\beta(x, \phi[n;x]) & = \beta_0 + \tfrac{1}{4}\phi[n;x]\\
\: & = \beta_{0}+ \tfrac{1}{4}\!\!\int_0^{\infty}\!\!K(x, x')n(x', t)\dd{x'},
\end{aligned}
\label{cannibalism_related}
\end{equation}
where $\beta_0, \mu$ are constants. Using Eq.~\eqref{cannibalism_related}
along with $K(x', x)=K_{1}(x',x) \equiv \theta(x'-2)\theta(2-x)$ in 
Eq.~\eqref{LVmodel}, we compute and plot the total steady-state
population at as a function of $\beta_0$ and $\mu$ in
Fig.~\ref{harvest}(b). When $\beta_{0}$ is fixed, the total population
is found to first increase with the death rate $\mu$ until
$\mu\rightarrow\beta_{0}^+$ when the population starts to diminish.
This implies that for the cannibalism-rate-dependent birth rate
$\beta$ defined in Eq.~\eqref{cannibalism_related}, overcompensation
can arise.

\subsection{Harvesting-induced overcompensation}
\label{app:harvesting}

Populations can also overcompensate selective harvesting
\citep{harvesting_palm_2003,WEIDEL,ZIPKIN} which we can model by
incorporating, as shown by Eq.~\eqref{LVmodel_harvest}, an
age-dependent harvesting term $h(n;x,t)$ that can be a nonlinear
function of $n$ \citep{diedrichs2019using}.
  
We explore age-dependent harvesting $h(n; x, t)$ that
preferentially removes older populations and show numerically that 
overcompensation can arise for the two forms of harvesting 

\begin{equation}
\begin{aligned}
h_1 = & \min\{n(x, t), h\}\theta(x-2), \\
h_2 = & \frac{hn(x, t)}{n(x, t)+n_{1/2}}\theta(x-2),
\end{aligned}
    \label{h_def}
\end{equation}
where $h$ is the intrinsic maximum harvesting rate and $n_{1/2}$ is a
constant half-saturation density.  Both effective harvesting rates
$h_1$ and $h_{2}$ vanish with the population densities $n(x,t)$,
saturate to $h$ when $n(x,t)\gg n_{1/2}$, and increase with the
parameter $h$.  We set all other dimensionless coefficients in
Eq.~\eqref{LVmodel_harvest} to $K(x',x) =K_{4}(x',
x)=(x'-x)\theta(x'-x),\,\beta=1,\, \mu=0.5,\, n_{1/2} = 1$.  In
Fig.~\eqref{harvest}(a), we plot the plot steady-state population
$N^*$ for scenarios $h_1$ and $h_2$ as a function of $h$.  The total
population $N^{*}$ is seen to increase with $h$ for both harvesting
strategies, indicating overcompensation in response to increased
harvesting rate.

\subsection{Overcompensation following changes in birth rate}
\label{birth_OC}

The usual ``hydra effect'' overcompensation is described by a steady
state total population that increases with the \textit{death} rate. In
all of our examples, the total steady-state population increased with
birth rate $\beta$.  One can show that if $K(x', x) \geq 0$ and $K(x',
x)=0$ for $x>X$ or $x'\leq x$, the steady-state solutions to
Eq.~\eqref{steady_state} that correspond to birth rates
$\beta_1(x)>\beta_2(x)$, $n_{\beta_1}^*(x)$ and $n_{\beta_2}^*(x)$,
are such that the total steady-state total populations

\begin{equation}
N^*_{\beta_1}\coloneqq \int_0^{\infty}\! n^*_{\beta_1}(x)\dd{x} 
> N^*_{\beta_2}\coloneqq \int_0^{\infty}\! n^*_{\beta_2}(x)\dd{x}.
\label{comparison}
\end{equation}
In fact, steady-state solution $n^*(x), x\geq 0$ to
 Eq.~\eqref{steady_state} can be expressed in terms of $n^*(X)$

\begin{equation}
\begin{aligned}
n^*(x\leq X) =  & n^*(X)e^{\int_x^X\mu(x')\dd{x'}} e^{\int_x^X \int_{x'}^{\infty}\!
  K(y, x')n^*(y)\dd{y}\dd{x'}},\\
n^*(x\geq X) = & n^*(X)e^{-\int_X^x\mu(x')\dd{x'}}.
\end{aligned}
\label{Eq_n_steady}
\end{equation}
We conclude from Eq.~\eqref{Eq_n_steady} that if
$n_{\beta_1}^*(X)>n_{\beta_2}^*(X)$ then
$n_{\beta_1}^*(x)>n_{\beta_2}^*(x), x\geq 0$ and therefore
Eq.~\eqref{comparison} still holds.

On the other hand, if $n_{\beta_1}^*(X)\leq n_{\beta_2}^*(X)$, we
conclude from Eq.~\eqref{Eq_n_steady} that $n_{\beta_1}^*(x)\leq
n_{\beta_2}^*(x), x\geq 0$. However, from Eq.~\eqref{eq:Euler-Lotka},
we have

\begin{equation}
\int_0^{\infty}\!\beta_i(x) e^{-\int_0^x \tilde{\mu}_i(a)\dd{a}}\dd{x}=1,\,\,\, i=1, 2
\end{equation}
where 

\begin{equation}
\begin{aligned}
\tilde{\mu}_1(x) & = 
\mu(x) +\!\int_x^{\infty}\!\!K(x', x)n^*_{\beta_1}(x')\dd{x'} 
\leq \tilde{\mu}_2(x) \\
\: & = \mu(x) +\!\int_x^{\infty}\!\!K(x', x)n^*_{\beta_2}(x')\dd{x'}.
\end{aligned}
\end{equation}
Therefore,

\begin{equation}
\begin{aligned}
&1=\int_0^{\infty}\!\beta_1(x)e^{-\int_0^x \tilde{\mu}_1(a)
\dd{a}}\dd{x} \\
&\qquad\qquad\quad > \int_0^{\infty}\!\beta_2(x) e^{-\int_0^x 
\tilde{\mu}_2(a)\dd{a}} \dd{x} = 1
\end{aligned}
\end{equation}
is a contradiction implying $n_{\beta_1}^*(x)> n_{\beta_2}^*(x), x\geq
0$ and that the total steady-state population $n_{\beta_1}^*(x)>
n_{\beta_2}^*(x), x\geq 0$ always increases with birth rate when
$K\geq 0$ and the predation is unidirectional (old-eat-young).

In scenarios in which the younger population can prey on the older
population, and $K$ can be negative, steady-state total populations
\textit{can} decrease with the birth rate, \textit{i.e.}, the
steady-state total population ``overcompensates" as the birth rate
decreases. As an example, we assume a dimensionless predation rate of
the form

\begin{equation}
\tilde{K}(x', x) \equiv 2\theta(X-x') -1, 
\end{equation}
set $X, \mu$, and $\beta$ to be dimensionless constants, and
investigate how the population varies with $\beta$.  Here, the young
population $x'<X$ suppresses the whole population as $\tilde{K}(x',
x)=1>0,\, x'<X$, while the old population $x'\geq X$ has a positive
effect on the whole population since $\tilde{K}(x', x)=-1,\, x'\geq
X$.  The explicit solution for the steady-state population is

\begin{equation}
\begin{aligned}
n^*(x) = & \frac{\beta (\beta-\mu)e^{-\beta x}}{\big(1-2e^{-\beta X}\big)},\\
N^*= & \!\int_0^{\infty}\!\!n^*(x)\dd{x} = \frac{(\beta-\mu)}{\big(1-2e^{-\beta X}\big)}.
\end{aligned}
\end{equation}
Upon taking the derivative $\partial_{\beta}N^{*}$, we find

\begin{equation}
\frac{\p N^*}{\p \beta} = \frac{1 -2e^{-\beta X} -2(\beta-\mu)Xe^{-\beta X}}
{\big(1-2e^{-\beta X}\big)^2},
\end{equation}
and specifically, $\big(\partial N^{*}/\partial \beta\big)<0$ if
$2+2(\beta-\mu)X > e^{\beta X}$. Therefore, if the interspecific
interaction $K$ allows younger individuals to suppress the overall
population, the steady-state population can overcompensate by
decreasing as the birth rate $\beta$ is increased.


\section{Analysis of the discretized ODE system Eq.~\eqref{pieceODE}}
\label{DISCRETE}

\subsection{Uniqueness of the positive equilibrium of the ODE Eq.~\eqref{pieceODE}}

We shall first show that there is at most one positive steady-state
solution $\{n_i^*\}$ of the discretized ODE Eq.~\eqref{pieceODE}. We
will prove by contradiction and assuming two distinct positive
equilibria.  The positive steady-state solution to
Eq.~\eqref{pieceODE}, if it exists, satisfies the backward difference
equation
%
\begin{align}
n_{i-1}^*  = & \Big(1 + \Delta{x}\mu_{i} + \Delta{x}\sum_{j \geq i} 
K_{j, i} n_{j}^* \Big) n_{i}^*, \,\,\, 1 <  i \leq L-1 \label{L_minus_two}\\
n_{L-1}^*  = & \Delta{x}\big(\mu_{L}+ K_{L,L} n_{L}^{*}\big) n_{L}^* \label{L_minus_one}\\
\sum_{i=1}^L\beta_in_i^* = & \mu_0n_{0}^* +
n_0^* \sum_{j=0}^{L}K_{j, i}n_j^* + \frac{n_0^*}{\Delta{x}}.
\label{ode_steady_state_solution}
\end{align}
%
We proceed by showing that if $\{m_i^*\}$ and $\{n_i^*\}$ are two
positive steady states, then $m_{L}^*=n_{L}^{*}$.  If
$m_{L}^*=n_{L}^*$, then by induction, $m_{i}^*\equiv n_{i}^*$. If
$n_{L}^{*}>m^{*}_{L}$, then $n_{L-1}^* > m_{L-1}^*$ by
Eq.~\eqref{L_minus_one}. Since $K_{i, j} \geq 0$, we observe that
$K_{L-1,L} n_{L}^* \geq K_{L-1,L} m_{L}^*$.  This inequality further
demonstrates that $n_{L-2}^* > m_{L-2}^*$ combined with
Eq.~\eqref{L_minus_two}. Thus, by induction, $n_{L}^{*}>m^{*}_{L}$ leads
to $n_{i}^{*}>m_{i}^{*}$ for all $i \in \left\{ 0,1,...,L \right\}$.

Next, let $n_i(t)$ be solutions to Eq.~\eqref{pieceODE} with the initial
value equal to the steady state $n_i^*$. Let $B(t) \coloneqq \sum_{i}
\beta_i n_i(t)$ be the newborn population at time $t$. Then, for any
compartment $i$, the population $n_i(t)$ at time $t$ is composed of
two parts: the survivors from the initial $t=0$ population and those
who were born in $(0, t)$. In order to characterize survival, we
define $\{s_{j,i}(t)\}_{j=0}^{L}$ to be the solution of

\begin{equation}
\begin{aligned}
\frac{\dd s_{i,0}(t)}{\dd t} = & - \mu_0 s_{i, 0}(t) - s_{i, 0}(t)\sum_{j=0}^{L}\! K_{j,0} n_j(t)
- \frac{s_{i,0}(t)}{\Delta{x}}, \\
\frac{\dd s_{i,j}(t)}{\dd t} = & - \mu_{i} s_{i,j}(t) - s_{i,j}(t)\sum_{\ell=j}^{L}\!K_{\ell,j}n_{\ell}(t)
\\
\: & \qquad\qquad+\frac{s_{i,j-1}(t)- s_{i,j}(t)}{\Delta{x}},\quad\,  0<j<L,\\
\frac{\dd s_{i,L}(t)}{\dd t} = & - \mu_{L}s_{i,L}(t) - s_{i,L}(t)
K_{L,L} n_{L}(t)+\frac{s_{i,L-1}(t)}{\Delta{x}}, \\
\end{aligned}
\label{pieceODE_survival}
\end{equation}
with the initial condition $s_{i,j}(0)=1, i=j$ and $s_{ik}(0)=0, ~\forall k\neq i$.

Note that an initial condition $n_{i}(0) = n_{i}^{*}$ implies that
$n_{i}(t) = n_{i}^{*}$.  Since death rates and interaction terms do
not explicitly depend on time, the survival fraction is
time-translation invariant, \textit{i.e.}, the survival from
compartment $i$ at time $t = 0$ to the compartment $j$ at time $t =
T-\tau$ is the same as survival from compartment $i$ at time $t =
\tau$ to the compartment j at time $t = T$.  Therefore, the solution
to Eq.~\eqref{pieceODE} can be written as

\begin{equation}
n_i(t) =  \sum_{j=0}^L n_{j}(0)s_{j, i}(t) +\!\int_0^t\!\!B(\tau) s_{0, i}(t-\tau)\dd{\tau},
\label{nB}
\end{equation}
where $B(t)\coloneqq \sum_{i=0}^L \beta_in_i(t)$ is the total birth
rate at time $t$.  Since every individual eventually dies, we have
$\lim_{t \rightarrow \infty}s_{i, j}(t) =0$ for all $i,j$. Therefore,
using the solution in Eq.~\eqref{nB} in $B(t)$, the birth rate can be
decomposed into contributions from the initial population and from the
population born within time $(0, t)$:
\begin{equation}
B(t) = \sum_{i,j=0}^{L} \beta_j n_i(0) s_{i, j}(t) +\!\int_0^t\!\!B(t - \tau) 
\sum_{i=0}^{L} \beta_is_{0, i}(\tau) \mathrm{d}\tau.
\end{equation}
At steady state, we introduce the lower and upper bounds of $B(t)$:
\begin{equation}
\begin{aligned}
&\int_0^t \!B(t-\tau) \sum_{i=0}^{L} \!\beta_i s_{0, i}(\tau) \dd{\tau}
\leq B(t)
\\ & \quad \leq \!\int_0^t \!B(t-\tau) \sum_{i=0}^{L}\beta_i s_{0, i}(\tau) \dd{\tau}
+ \!\sum_{i,j=0}^{L} n^{*}_{i} \max_i\{\beta_i\} s_{i, j}(t),
\end{aligned}
\label{birthbalance}
\end{equation}
where the left-hand side represents the birth rate at time $t$
generated by individuals born with $(0, t)$ and the right-hand side
are birth rate of newborns at time $t$ that are offspring of
individuals born within $(0, t)$, plus the maximum possible number of
offspring that the initial population could give birth to at time $t$.
When $n_i(0)=n_i^*$, $B(t)\coloneqq B$ is a constant and the limit
$t\rightarrow\infty$ forces the lower and upper bounds to converge
yielding

\begin{equation}
B = B\!\int_0^{\infty}\!\sum_{i=0}^{L} \beta_i s_{0, i}(\tau) \dd{\tau},
\end{equation}
which is the discrete analogue of the $R_{0}=1$ condition of
Eq.~\eqref{eq:Euler-Lotka} where the factor $\int_0^{\infty}
\sum_{i=0}^{L} \beta_i s_{0, i}(\tau) \dd{\tau}$ on the
right-hand side is the expected offspring that one individual has
during its lifetime.

Similar to the proof of uniqueness in Appendix \ref{UNIQUE}, it is
intuitively clear that $s_{0, i}(a)$ as well as $\int_0^{\infty}
\sum_{i=0}^{L} \beta_i s_{0, i}(\tau) \dd{\tau}$ monotonically
decreases with with increasing effective death rate
$\tilde{\mu}_i$. We demonstrate this by explicitly computing

\begin{equation}
\begin{aligned}
1 & =\int_0^{\infty}\!\sum_{i=0}^{L}\beta_i s_{0, i}(\tau) \dd{\tau} \\
\:& = \sum_{i=1}^{L-1}\prod_{j=0}^{i-1}\frac{1}{(1+\Delta{x}\tilde{\mu}_j)} \int_0^\infty\!\!\!
\big(\tilde{\mu}_i+\tfrac{1}{\Delta{x}}\big)\beta_{i} t e^{-(\tilde{\mu}_i+1/\Delta{x})t}\dd{t} \\
\: & \qquad\qquad  + \prod_{j=0}^{L-1}\frac{1}{(1+\Delta{x}\tilde{\mu}_j)} 
\int_0^\infty \mu_{L}\beta_{L} t e^{-\tilde{\mu}_{L}t}\dd{t},\\
\: & =\sum_{i=1}^{L-1}\prod_{j=0}^{i-1}\frac{\beta_{i}\Delta x}
{(1+\Delta{x}\tilde{\mu}_j)(\Delta{x}\tilde{\mu}_i+1)} + \prod_{j=0}^{L-1}
\frac{\beta_{L}}{\mu_{L}(1+\tilde{\mu}_j)},
\end{aligned}
\label{expected}
\end{equation} 
where $\tilde{\mu}_i\coloneqq \mu_i+\sum_{j>i}K_{j, i}n_j^*$. The
first term on the right-hand side of Eq.~\eqref{expected} is the
summation of the expected number of offspring that an individual gives
birth to while in the $i^{\text{th}},\, i< L$ stage multiplied by the
probability that it will survive until the $i^{\text{th}}$ stage.  The
second term on the right-hand side is the expected number of offspring
that an individual gives birth while in the $L^{\text{th}}$ stage
multiplied by the probability that it survives to the $L^{\text{th}}$
stage. If $n_i^*>m_i^*,\, i=0,\ldots,L$ and there exists at least one
$K_{j, i}>0$, then $\tilde{\mu}^n_i > \tilde{\mu}^m_i$ and
Eq.~\eqref{expected} cannot be satisfied by two distinct steady-state
solutions $\{m_i^*\} \neq \{n_i^*\}$.

\subsection{Permanent overcompensation is precluded in
  two-compartment ODE models}

In the following discussion, we will exclude the artificial
self-inhibition term $K_{i,i}$ as a result of binning the age
structure into a finite number of compartments. We start by
considering the simplest two-compartment model by imposing some
additional assumptions on the coefficients. Setting $L=1$ (two
compartments) in Eq.~\eqref{pieceODE}, we find

\begin{equation}
\begin{aligned}
&\frac{\dd n_0}{\dd t}  = -\mu_0 n_0 - K_{1, 0} n_1 n_0 
+ \beta_1 n_1 - \frac{n_0}{\Delta{x}}, \,\,\,\\
&\frac{\dd n_1}{\dd t}  = -\mu_1 n_1 +\frac{n_0}{\Delta{x}}.
\end{aligned}
\label{ODEmodel1}
\end{equation}
Eq.~\eqref{ODEmodel1} admits a unique steady state at
 
\begin{equation}
    (n_0^*, n_1^*)=\bigg(\frac{\beta_1 - \mu_1-\mu_0\mu_1\Delta{x}}{K_{1, 0}},
\frac{\beta_1 - \mu_1-\mu_0\mu_1\Delta{x}}{K_{1, 0}\mu_1\Delta{x}}\bigg),
\end{equation}
which, as is the total population $n_{0}^{*}+n_{1}^{*}$, monotonically
decreasing with either $\mu_0$ or $\mu_1$, indicating that
steady-state overcompensation cannot arise. The Jacobian matrix at the
fixed point is

\begin{equation}
{\bf J} =\left[
\begin{array}{cc}
\displaystyle -\mu_0 - K_{1, 0} n_1 -\tfrac{1}{\Delta{x}} & 
\displaystyle -K_{1, 0}n_0+\beta_1\\[8pt]
\displaystyle \tfrac{1}{\Delta{x}} & \displaystyle -\mu_{1}
\end{array}
\right].
\end{equation}
which has two negative eigenvalues if the equilibrium $(n_0^*,
n_1^*)>0$. Therefore, the steady state is stable and we do not expect
periodic oscillations in a small neighborhood around this fixed point.
Note that limit cycles may still exist which is not dependent on the
stability of the positive equilibrium and is difficult to directly
prove. Note that the oscillations demonstrated for a two-compartment
model studied in \citep{Hastings_Costantino1987} arose from a
different form of the birth rate, which we keep constant.

\subsection{Undamped oscillations are precluded in a three-compartment model}

Setting $L=2$ in Eq.~\eqref{pieceODE}, we obtain
\begin{equation}
    \begin{aligned}
\frac{\dd n_0}{\dd t} & = -\frac{n_0}{\Delta{x}} - \mu_{0} n_{0}-
K_{1, 0}n_{1}n_{0}-K_{2, 0}n_{2}n_{0}+\beta_1 n_1 + \beta_2 n_2,\\
  \frac{\dd n_1}{\dd t} & = -\frac{n_1}{\Delta{x}}+\frac{n_0}{\Delta{x}}
-\mu_{1}n_{1} - K_{2, 1}n_{2}n_{1},\\
  \frac{\dd n_2}{\dd t} & = \frac{n_1}{\Delta{x}} -\mu_{2}n_{2}.
    \end{aligned}
    \label{ODE3D}
\end{equation}
First, we demonstrate that this three-compartment model can exhibit
overcompensation by considering a simple specific set of parameters:
$\beta_2= \beta_2' + \mu_2$, $\beta_2' \geq 0$, $\mu_1=0$,
$K_{1,0}=K_{2,1}=0$. Equations~\eqref{ODE3D} then simplify to
\begin{equation}
\begin{aligned}
    \frac{\dd n_0}{\dd t} & = -\frac{n_0}{\Delta{x}} - K_{2, 0}n_2n_0 
+ (\beta_2'+\mu_2)n_2+\beta_{1} n_{1}, \\
    \frac{\dd n_1}{\dd t} & = \frac{n_0-n_1}{\Delta{x}}, \\
    \frac{\dd n_2}{\dd t} & = \frac{n_1}{\Delta{x}}-\mu_2 n_2
    \label{3dmodel}
\end{aligned}
\end{equation}
which admits the positive steady state

\begin{equation}
    (n_0^*, n_1^*, n_2^*) =\left(\frac{\beta_1\mu_2\Delta{x}+\beta_2'}{K_{2,
    0}},\frac{\beta_1\mu_2\Delta{x}+\beta_2'}{K_{2, 0}},
\frac{\beta_1\mu_2\Delta{x}+\beta_2'}{\Delta{x}\mu_2K_{2, 0}}\right)
\end{equation}
and the total steady-state population 

\begin{equation}
    N(\mu_2)\coloneqq n_0^*+n_1^*+n_2^* 
=2\frac{\beta_1\mu_2\Delta{x}+\beta_2'}{K_{2, 0}} 
+ \frac{\beta_1\mu_2\Delta{x}+\beta_2'}{\Delta{x}\mu_2K_{2, 0}}.
\end{equation}
Therefore, $\partial N(\mu_2)/\partial\mu_{2} =
\frac{2\beta_1\Delta{x}}{K_{2, 0}} - \frac{\beta_2'}{K_{2,
    0}\mu_2^2\Delta{x}}$ indicates that the total population at
equilibrium $N(\mu_2)$ will increase with the death rate of the oldest
population $\mu_2$ if $\mu_2 >
\sqrt{\frac{\beta_2'}{2\beta_1\Delta{x}^2}}$. So in order to observe
overcompensation, at least three compartments are needed.

Next, we show that the positive steady state of the three-compartment
model Eqs.~\eqref{ODE3D}, if it exists, is stable. This statement
holds for general parameter values in Eqs.~\eqref{ODE3D}. The
steady-state populations $n_{i}^{*}$ obey the relationships
\begin{equation}
    \begin{aligned}
n_{1}^*  = & \Delta{x}\mu_{2}n_{2}^*, \quad\\
n_{0}^* = & \big(\Delta{x}+\mu_1\Delta{x}^2+\Delta{x}^2K_{2, 1}n_2^*\big) \mu_{2} n_{2}^*,\\
\Delta{x}\,\beta_1\mu_2 + \beta_2  = & \mu_{2}\Big(\mu_0+\Delta{x}K_{1, 0}
    \mu_{2}n_{2}^{*}+K_{2, 0}n_{2}^{*}+\tfrac{1}{\Delta{x}}\Big)\\
\: &\qquad\quad\times\big(\Delta{x}+\Delta{x}^2\mu_{1} + \Delta{x}^{2} K_{2, 1}n_{2}^{*}\big).
    \end{aligned}\label{eq:ss_constraints}
\end{equation}
Therefore, this system contains one real positive fixed point whenever
a positive root for $n_{2}^{*}$ satisfies the last equation in
\eqref{eq:ss_constraints}. This occurs for parameters values for which
 
\begin{equation}
\beta_1\mu_2\Delta x + \beta_{2} > \mu_{2}\big(\mu_0
  \Delta{x}+1\big) \big(\mu_{1}\Delta{x}+1\big).
\label{root_condition}
\end{equation}
The Jacobian matrix at this fixed point is

\begin{strip}
\begin{equation}
{\bf J} =\left[
\begin{array}{ccc}
\displaystyle -\mu_0 - K_{1, 0} n_{1}^{*} -\tfrac{1}{\Delta{x}}-K_{2, 0}n_2^* & 
\displaystyle -K_{1, 0}n_0^*+\beta_1  &  -K_{2, 0}n_0^*+\beta_2\\[8pt]
\displaystyle \tfrac{1}{\Delta{x}}  & \displaystyle -\tfrac{1}{\Delta{x}}
-\mu_{2}-K_{2, 1}n_{2}^{*} &-K_{2, 1}n_{1}^{*}\\[8pt]
0 & \displaystyle \tfrac{1}{\Delta{x}} & -\mu_2
\end{array}
\right]
\end{equation}
whose eigenpolynomial  is

\begin{equation}
\begin{aligned}
f(\lambda) \equiv \text{det}(\lambda \mathds{I}-{\bf J})= & 
\big(\lambda+\mu_0 + K_{1, 0} n_1^* +\tfrac{1}{\Delta{x}}
+K_{2, 0}n_{2}^{*}\big)\big(\lambda+\tfrac{1}{\Delta{x}}+\mu_1
+K_{2, 1}n_{2}^{*}\big)\big(\lambda+\mu_2)+ \tfrac{1}{\Delta{x}^2}(K_{2, 0}n_0^*-\beta_2\big)\\
\: & \qquad\qquad\qquad\qquad + \tfrac{1}{\Delta{x}}K_{2, 1}n_1^* \big(\lambda + \mu_0 + K_{1, 0} 
n_1^* +\tfrac{1}{\Delta{x}}+K_{2, 0}n_{2}^{*} \big)+ 
\tfrac{1}{\Delta{x}}\big(\lambda+\mu_2)(K_{1, 0}n_0^*-\beta_1\big),
    \end{aligned}
\end{equation}
\end{strip}
\noindent where $\mathds{I}$ is the identity matrix. To simplify this
expression, define the effective death rates to be $\tilde \mu_0 =
\mu_0 + n_1^* K_{1,0} + n_2^*K_{2,0}$ and $\tilde{\mu}_1 = \mu_1
+n_2^* K_{2,1}$. Then, the eigenpolynomial can be
expressed as
%
\begin{align}
\begin{split}
f(\lambda) = & \lambda^3 + C_2 \lambda^2 + C_1 \lambda + C_0, \\
C_2 & = \mu_2+\tilde{\mu}_0+\tilde{\mu}_1+\tfrac{2}{\Delta x} \\
C_1 & = K_{1,0} n_1^* \big( \tilde{\mu}_1 + \tfrac{1}{\Delta x}\big) 
+\!K_{2,1} n_2^* \mu_2+\mu_2 C_2+\tfrac{\beta_2}{\Delta x^2 \mu_2}\\
C_0 & = \tfrac{\mu_2}{\Delta x} \big(K_{1,0} n_1^* \Delta x 
\tilde{\mu}_1+K_{1,0} n_1^*+K_{2,0} n_2^* \Delta x \tilde{\mu}_1
\\
\: & \hspace{2cm} +K_{2,0} n_2^*+K_{2,1} n_2^* \Delta x \tilde{\mu}_0+K_{2,1} n_2^*\big)
\end{split}
\end{align}
Here, we have employed Eq.~\eqref{eq:ss_constraints} to replace
$n_0^*$, $n_1^*$, and $\beta_1$ by simple terms involving $n_2^*$.
Note that our parameters are all non-negative.  We may reasonably
parameterize our model such that at least one $K_{i,j}>0$, at least
one $\mu_i >0$, at least one $\beta_i >0$, and all $n_i^* >0$. Under
such assumptions, $C_0, C_1, C_2 > 0$. Then, $f(\lambda)$ is
monotonically increasing on $(0, + \infty)$. Therefore, $f(\lambda)$
has no positive real root.

What remains is to show that $f(\lambda)$ cannot have a pair of
complex roots with positive real parts, which we prove by
contradiction. Suppose such a pair of complex roots $\lambda_\pm$
exists with $\operatorname{Re}(\lambda_\pm)>0$. Recall that
$f(\lambda)$ is a polynomial of degree 3, and, from our discussion
above, has a negative real root $\lambda_1$.  $f(\lambda)$ can be
factorized as $f(\lambda)= (\lambda - \lambda_1)(\lambda^2 + b \lambda
+ c)$ with $b = -(\lambda_+ + \lambda_-) <0$, $\lambda_1 <0$, and
$-\lambda_1 + b = C_2$. Thus, $\lambda_1 < - C_2$ and because
$f(\lambda)$ is monotonically increasing in $(- \infty, - C_2)$,
$f(-C_2) \geq f(\lambda_1)=0$.

We next demonstrate the polynomial $-f(-\lambda - C_2)=\lambda^3 + D_2
\lambda^2 + D_1 \lambda^1 + D_0$ is monotonically increasing for
$\lambda > 0$ and that $D_0 > 0$.  Through straightforward algebra, we
note that $D_2 = 2 C_1> 0$, $D_1 = C_2^2 + C_1 > 0$, $D_0 = C_1 C_2 -
C_0 >0$. To see $D_0 > 0$, we just claim that every term in $C_0$ can
be written as one term in the product $C_1 C_2$. For example, the
first term $K_{1,0} n_1^* \tilde{\mu}_1 \mu_2$ in $C_0$ can be written
as the product of the $K_{1,0} n_1^* \tilde{\mu}_1$ term from $C_1$
and the $\mu_2$ term from $C_2$. Also note that every term in $C_1$
and $C_2$ is positive. Thus, we have $D_0 > 0$. We conclude that
$f(-C_2) < 0$. This contradicts $f(-C_2) \geq f(\lambda_1)=0$ and
precludes complex roots $\lambda_\pm$ with positive real parts.

Combining previous uniqueness statement and stability analysis, the
system Eq.~\eqref{ODE3D} admits at most one positive steady state
which must be stable. Therefore, a three-compartment model precludes
oscillatory solutions in a close neighborhood around the steady state
in the total population since the positive steady state is stable.

\subsection{Higher-order reduced ODE models}
\begin{figure*}[htb]
    \centering
    \includegraphics[width=6.8in]{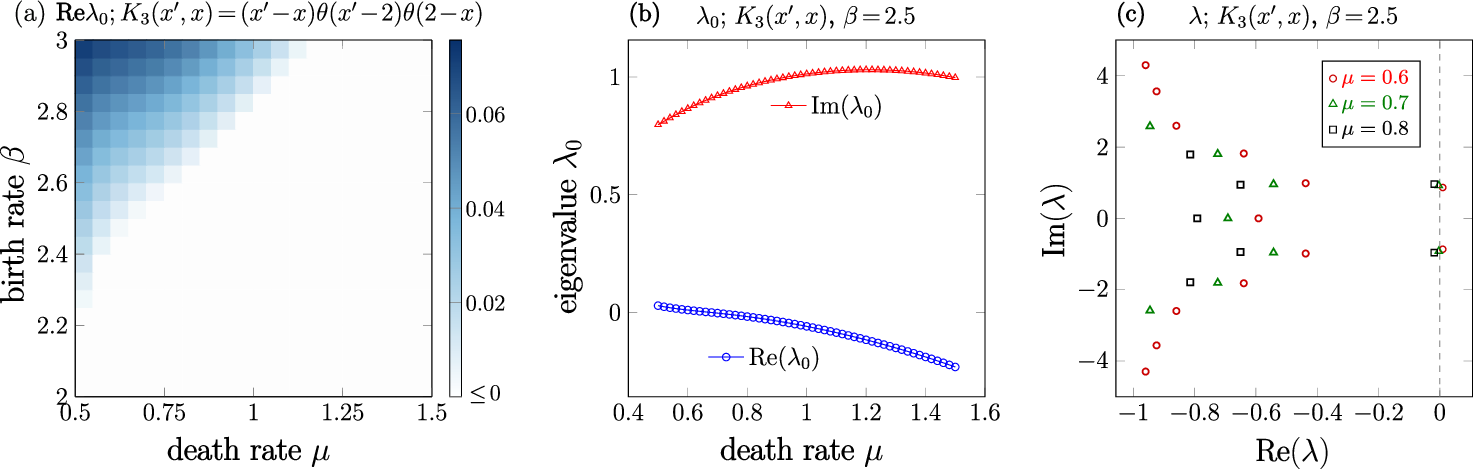}
    \caption{\small (a) Heatmap of the real part of the principle
      eigenvalue $\lambda_0$ associated with the Jacobian matrix of
      the discretized, 500 ODE system Eq.~\eqref{pieceODE} (with
      $L=499$) at its fixed point. The top left region takes positive
      real values. (b) Dependence of the largest eigenvalue
      $\lambda_0$ on $\mu$ for $\beta=2.5$. When $\mu$ is small,
      $\text{Re}\lambda_0>0$, which indicates an unstable positive
      equilibrium.  In (a) and (b), $\beta,\mu$ are age-independent,
      and the cannibalism rate is derived from $K_{3}$ in
      Eq.~\eqref{cannibalism_rate_oscillation}. (c) The first five
      eigenvalues for $\mu=0.6, 0.7, 0.8$ (open circle, triangle,
      square, respectively). When $\mu=0.6$, $\lambda_0$ has a
      positive real part; when $\mu=0.7, 0.8$, $\lambda_0$ has a
      negative real part, implying stability of the steady
      state. \label{fig:stability}}
\end{figure*}
For higher-order ODE models with $L+1, L\geq2$ compartments, we can
consider the special case

\begin{equation}
    \begin{aligned}
\frac{\dd n_0}{\dd t} = & -\frac{n_0}{\Delta{x}}
      - K_{L, 0}n_0n_{L} + \beta_1n_1+(\beta_{L}'+\mu_{L})n_{L},\\
\frac{\dd n_i}{\dd t} = & \frac{n_{i-1}-n_i}{\Delta{x}},~i=1,2...,L-1,\\
\frac{\dd n_{L}}{\dd t} = & \frac{n_{L-1}}{\Delta{x}}-\mu_{L} n_{L}.
    \end{aligned}
    \label{ndmodel}
\end{equation}
which has the equilibrium 

\begin{equation}
\begin{aligned}
n_i^* = & \frac{\beta_1\mu_{L}\Delta{x}+\beta_{L}'}{K_{L, 0}},\,\,\, i=0,...,L-1; \\
n_{L}^* = & \frac{\beta_1\mu_{L}\Delta{x}+\beta_{L}'}{K_{L, 0}\mu_{L}\Delta{x}}.
\end{aligned}
\end{equation}
The total population at equilibrium as a function of $\mu_{L}$ is
$N^*(\mu_{L})\coloneqq \sum_{i=0}^{L} n_i^* =
L\frac{\beta_1\mu_{L}\Delta{x}+\beta_{L}'}{K_{L, 0}} +
\frac{\beta_1\mu_{L}\Delta{x}+\beta_{L}'}{K_{L,
    0}\mu_{L}\Delta{x}}$. Therefore, $\frac{\dd N(\mu_{L})}{\dd
  \mu_{L}} = \frac{L\beta_1\Delta{x}}{K_{L, 0}} - \frac{\beta_{L}'}{
  K_{L, 0}\mu_{L}^2\Delta{x}}$, indicating that the total population
at equilibrium is increasing with $\mu_{L}$ as long as
$\mu_{L}>\sqrt{\frac{L\beta_{L}'}{\beta_{1}\Delta{x}^2}}$.  Thus, for
higher-order compartment ODE models, overcompensation of the total
equilibrium population to increases in death rate of certain
subpopulations is always possible.

The equilibrium of multi-compartment ODE models, if it exists, could
be unstable. We now switch back to the model discussed in the main
text in Section \ref{example2}. The numerical solution of the
structured population obtained by the finite volume method, which is a
500-compartment ODE Eq.~\eqref{pieceODE} with $L=499$ displays
undamped oscillatory behavior. We numerically analyzed the stability
of the positive equilibrium of the PIDE Eq.~\eqref{LVmodel} with the
{cannibalism} rate $K(x',x)$ defined by
Eq.~\eqref{cannibalism_rate_oscillation} and the same age-independent
birth rate $\beta$ and death rate $\mu$ as used in
subsection~\ref{example2}.  As a surrogate of the PIDE
Eq.~\eqref{LVmodel}, we numerically analyzed the derived ODE system
Eq.~\eqref{pieceODE} in subsection~\ref{example2} with
$\mathrm{d}x=0.02,~ L=499$. In Eqs.~\eqref{L_minus_two} and
\eqref{L_minus_one}, $n^*_{i-1}$ is completely determined by $\left\{
n_j^*:j \geq i \right\}$. Therefore, the steady-state solution $n^*_i,
i=0,..,L-1$ can be parameterized by the value of $n^*_{L}$,
\textit{i.e.}, $n_i^* = n_i^*(n^*_{L})$.  Considering the newborn
individuals, we employed the bisection method to find a proper
positive $n^*_{L}$ such that Eq.~\eqref{ode_steady_state_solution} is
satisfied.

We then consider the Jacobian matrix ${\bf J}(n^*)$ of the dynamical
system at the steady state and numerically find its eigenvalues. We
denote the principle eigenvalue of ${\bf J}(n^*)$ with the largest
real part by $\lambda_0$. The eigenvector corresponding to $\lambda_0$
decays (grows) slowest for $\operatorname{Re}\lambda_0 <0$
($\operatorname{Re}\lambda_0>0$) and characterizes the long-term local
dynamical behavior of the system. Near the steady state, we found
that, corresponding to the region of oscillation described in
Fig.~\ref{oscillatory}(f), there is also a region of linearly unstable
steady states with $\textrm{Re} \lambda_0 >0$ shown in
Fig.~\ref{fig:stability}(a).

To better understand the correspondence between oscillation and
unstable steady states, we examined real and imaginary parts of
$\lambda_0$ as a function of $\beta$ in detail, as shown in
Fig.~\ref{fig:stability}(b). When $\beta=2.5$ is fixed, the real part
of the principal eigenvalue $\operatorname{Re}\lambda_0$ increases as
$\mu$ is decreased, vanishing at about $\mu\approx 0.7$.  At this
point $\lambda_0$ (and $\lambda_{0}^{*}$) become purely imaginary,
indicative of a Hopf-type bifurcation.  As $\mu$ is further decreased,
$\lambda_0$ and $\lambda_{0}^{*}$ acquire positive real parts. This
regime corresponds to the numerical result plotted in
Fig.~\ref{oscillatory}(c.d) where undamped oscillations are found to
arise when $\beta=2.5,\mu \leq 0.7$.

Generalizing to more compartments, if the Jacobian matrix ${\bf
  J}_{L}$ of the positive equilibrium $(n_0^*,\ldots,n_{L}^*)$ of the
$(L+1)$-compartment reduced ODE model Eq.~\eqref{pieceODE} has an
unstable equilibrium, we can assume that

\begin{equation}
\textbf{v}_{L}\in\mathbb{R}^{L+1},\,\, {\bf J}_{L}\textbf{v}_{L}
  =\lambda \mathds{I}\textbf{v}_{L}=(v_1,...,v_{L})\neq 0,\,\, \text{Re}\lambda>0.
\end{equation}
For $L'>L$, we can consider the following ODE model

\begin{equation}
\begin{aligned}
\frac{\dd n_0}{\dd t} = & - \mu_0(t)n_0(t) - n_0(t)\sum_{j=i}^{L}K_{j, 0}(t)n_j(t)
+\sum_{j=1}^{L}\beta_j(t)n_j(t) - \frac{n_{0}(t)}{\Delta{x}},\\
\frac{\dd n_i}{\dd t} = & - \mu_i(t)n_i(t)
  - n_i(t)\sum_{j=i}^{L}K_{j, i}(t)n_j(t)\\
\:  &\hspace{2.8cm} -\frac{n_i(t) -n_{i-1}(t)}{\Delta{x}},\quad L\geq i > 0,\\
\frac{\dd n_i}{\dd t} = & \frac{n_{i-1}(t)-n_i(t)}{\Delta{x}},\quad\,\,  i>L
\end{aligned}
\label{new_ODE}
\end{equation}
which has a positive equilibrium $(n_0^*,...,n_{L}^*,
n_{L+1}^*,...,n_{L'}^*), n_{i}^*=n_{L}^*, i>L$. Denoting the Jacobian
matrix of the equilibrium of the ODE Eq.~\eqref{new_ODE} to be
${\bf J}_{L'}$, it is obvious that $\lambda$ is also an eigenvalue of
${\bf J}_{L'}$ with the corresponding eigenvector

\begin{equation}
\textbf{v}_{L'}=\Big(v_1,...,v_{L},
\tfrac{1}{1+\Delta{x}\lambda}v_{L},\dots,\big(\tfrac{1}{1+\Delta{x}\lambda}\big)^{L'-L}v_{L}\Big).
\end{equation}
Therefore, all reduced ODE systems with $L'>L$ compartments have a
positive equilibrium whose Jacobian matrix has a positive eigenvalue
and the positive equilibrium can be unstable.

\end{document}